\documentclass[sigplan,nonacm]{acmart}

\usepackage{graphicx}
\usepackage{url}
\usepackage{amsmath}
\usepackage{xcolor}
\usepackage[linesnumbered,ruled,vlined]{algorithm2e}
\SetVlineSkip{1pt}
\usepackage{siunitx}
\usepackage{tabularx}
\usepackage{array}
\usepackage{mathtools}
\usepackage{braket}
\usepackage{booktabs}
\usepackage{multirow}
\usepackage{enumitem}

\newcommand{\secref}[1]{Sec.~\ref{#1}}
\newcommand{\figref}[1]{Fig.~\ref{#1}}

\makeatletter
\def\@ACM@checkaffil{}
\makeatother

\begin{document}

\title[]{MagiCFirm: A Runtime for Magic-State Cultivation with Algorithm–Hardware Co-Design}

\begin{abstract}
Magic-state cultivation offers a promising alternative for lowering the cost of non-Clifford operations in fault-tolerant quantum computing (FTQC). However, realizing cultivation in practice exposes two challenges: \textit{i)} the lack of an open-source classical runtime layer between logical software and physical control, and \textit{ii)} the latency constraints on protocol-specific decisions that determine magic-state readiness.
To address these challenges, we adopt an algorithm–hardware co-design approach. At the algorithm level, we develop a two-stage early-escape scheme that identifies an informative subset of detectors offline, constructs a compact decoding problem, and performs partial decoding in parallel with complete decoding at runtime, allowing high-confidence attempts to advance before full decoding completes. At the hardware level, we present {MagiCFirm}, a configurable runtime that combines offline-compiled microprograms with dedicated datapaths for detector construction, event processing, and protocol control, enabling end-to-end execution of magic-state cultivation. Across evaluated configurations, MagiCFirm reduces wall-clock magic-state preparation time by up to \(39.3\%\) at matched logical error rate. For a representative magic-state-bound workload, this translates to an estimated \(11\%\) reduction in overall application runtime.
\end{abstract}

\author{%
Jubo Xu\textsuperscript{1},
Abbas B. Ziad\textsuperscript{1},
Prakash Murali\textsuperscript{2},
and Hongxiang Fan\textsuperscript{1}%
}

\affiliation{%
  \institution{%
    \textsuperscript{1}Imperial College London
    \qquad
    \textsuperscript{2}University of Cambridge%
  }
}

% \email{%
% jubo.xu20@imperial.ac.uk,
% abbas.ziad25@imperial.ac.uk,
% pm830@cam.ac.uk,
% hongxiang.fan@imperial.ac.uk%
% }

% \author{Jubo Xu}
% \email{jubo.xu20@imperial.ac.uk}
% \orcid{0009-0009-6314-5833}
% \affiliation{%
%   \institution{Imperial College London}
%   \city{London}
%   \country{UK}
% }

% \author{Abbas B. Ziad}
% \email{abbas.ziad25@imperial.ac.uk}
% \orcid{0009-0008-2517-0619}
% \affiliation{%
%   \institution{Imperial College London}
%   \city{London}
%   \country{UK}
% }

% \author{Prakash Murali}
% \email{pm830@cam.ac.uk}
% \orcid{}
% \affiliation{%
%   \institution{University of Cambridge}
%   \city{Cambridge}
%   \country{UK}
% }

% \author{Hongxiang Fan}
% \email{hongxiang.fan@imperial.ac.uk}
% \orcid{0000-0003-2387-5611}
% \affiliation{%
%   \institution{Imperial College London}
%   \city{London}
%   \country{UK}
% }

\maketitle

\makeatletter
\def\@shortauthors{}
\makeatother

\section{Introduction}
\label{sec:introduction}
Large-scale quantum computation requires fault-tolerant quantum computing (FTQC), where quantum error correction (QEC) protects logical operations from physical faults~\cite{aharonov1999faulttolerantquantumcomputationconstant}. Universal FTQC further relies on low-error magic states~\cite{gottesman1998heisenbergrepresentationquantumcomputers}, whose preparation can consume substantial system resources and execution time~\cite{9251988}. Magic-State-Cultivation~\cite{gidney2024magicstatecultivationgrowing} has recently emerged as a resource-efficient preparation approach that improves a single encoded state in place through physical operations and post-selection. Its practical execution, however, extends beyond the quantum circuit itself: measurements must be interpreted, failed attempts restarted, and surviving states qualified through decoding before becoming available to downstream computation.

Existing high-level FTQC tools typically encapsulate this execution as a coarse-grained logical operation with an aggregate space--time cost~\cite{10.1145/3720416,10.1145/3695053.3730991,pflieger2026harvestresourceawarequantumcompilation,beverland2022assessingrequirementsscalepractical}, leaving its online realization largely unspecified. 
Bridging this abstraction gap exposes two challenges.
First, magic-state cultivation requires an execution layer that translates a logical preparation request into the sequence of physical operations, measurements, classical processing, and control-flow decisions that implement the protocol. Second, cultivation should execute under certain timing constraints. Runtime decisions depend on protocol-defined detectors constructed from noisy measurement histories and on classical processing, including real-time decoding. Because the encoded state must remain protected through continued QEC rounds, these decisions need to be produced within tight latency bounds rather than deferred arbitrarily. Moreover, stochastic measurement outcomes and decoding-dependent retries make cultivation's execution path variable, directly coupling classical processing latency to quantum execution time and resource occupancy. Together, these challenges call for an online execution layer that realizes cultivation as a concrete quantum--classical process while exposing its latency, retry behavior, and resource consumption to compilation and scheduling.

% Supporting this hidden process calls for an execution layer between logical software and physical control. 
% Classical digital processors bridge a similar abstraction gap through microprogrammed runtime control, where software-visible instructions are realized as sequences of low-level, hardware-specific control operations.
% FTQC protocols require analogous machinery, but under fundamentally different constraints: control decisions require constructing protocol-defined detectors from noisy measurement histories and interpreting the resulting events through classical processing, including real-time decoding. 
% Because the quantum state cannot be stalled and must remain protected through continued QEC rounds, this decision path must meet strict timing constraints. The resulting execution path and completion can also be stochastic, coupling classical processing latency to quantum runtime and resource provisioning. Together, these constraints expose a broader architectural problem of providing an online execution layer for FTQC protocols while making its classical latency, retry behavior, and resource cost explicit for compilation, scheduling, and resource estimation. Cultivation provides a concrete starting point, where a single \textsc{PrepareMagicState} request expands into a detector-driven quantum--classical process with stochastic, decoding-dependent completion.

To address these challenges, we first conduct a detailed performance profile and latency breakdown of magic-state cultivation to identify its key execution bottlenecks~(\secref{sec:background-motivation-and-formalism:motivation}). Guided by these observations and insights, we address the first challenge with \textbf{MagiCFirm}, a configurable hardware runtime that realizes magic-state cultivation end to end. 
The runtime couples offline-compiled microprograms for protocol sequencing with dedicated, configurable datapaths for detector construction and event processing, achieving sub-microsecond control with a lightweight hardware footprint. 
Within this execution path, decoding-based acceptance lies directly on the critical path to state availability. 
To address the second challenge of reducing classical processing latency, we further introduce a two-stage early-escape scheme that constructs a compact decoding problem offline and runs partial decoding over it alongside complete decoding at runtime, allowing high-confidence attempts to advance early and reducing preparation time by up to \(39.3\%\) at matched logical error rate. At a representative operating point where magic-state supply accounts for \(90\%\) of application execution time, the scheme yields an estimated \(11\%\) reduction in overall runtime.

In summary, this work makes the following contributions:

\begin{itemize}[leftmargin=*]
    \item A configurable \textbf{\emph{hardware runtime}} for magic state cultivation that combines offline-compiled microprograms with dedicated detector-construction and event-processing datapaths for low-latency execution across monolithic and distributed control systems~(\secref{sec:magic firm sys:controlsys}).

    \item A \textbf{\emph{partial gap estimation algorithm}} that uses Logical Ambiguity Profiling and Structural Closure to identify informative detectors, followed by graph-based DEM contraction to produce a compact decoding problem~(\secref{sec:magic firm sys:partialgap}).

    \item A \textbf{\emph{two-stage early-escape scheme}} that runs partial and complete decoding in parallel, advancing high-confidence attempts early and otherwise retaining complete decoding~(\secref{sec:magic firm sys:2stage}).
\end{itemize}

% \begin{itemize}[leftmargin=*]
% \item \textbf{Configurable cultivation control.}
% We develop an offline-compiled hardware runtime that realizes cultivation from a logical preparation request to state availability across monolithic and distributed physical-control systems.

% \item \textbf{Partial complementary-gap estimation.}
% We introduce Logical Ambiguity Profiling and Structural Closure to identify informative detectors, together with graph-based DEM contraction to construct a compact partial decoding problem.

% \item \textbf{Two-stage early escape.}
% We develop a guarded partial/complete decoding scheme that advances high-confidence attempts early and otherwise retains complete-gap evaluation, and characterize its end-to-end operating region and downstream impact.

% \end{itemize}
\section{Background and Motivation}
\label{sec:background-and-motivation}

\subsection{Background}
\label{sec:background-and-motivation:background}

\subsubsection{Quantum Error Correction}
\label{sec:background-and-motivation:background:qec}

QEC encodes logical qubits into redundant physical blocks and detects errors through stabilizer measurements. Major families include surface~\cite{fowler2012surface}, color~\cite{Bombin_2006}, and high-rate qLDPC codes~\cite{Breuckmann_2021}. For universality, Clifford gates are supplemented with the non-Clifford \(T\) gate, which is commonly implemented by gate teleportation consuming an encoded magic state, \(\lvert T\rangle_L=T_L\lvert+\rangle_L=(\lvert0\rangle_L+e^{i\pi/4}\lvert1\rangle_L)/\sqrt{2}\), making low-error magic-state preparation essential. Such states are primarily prepared through distillation~\cite{Litinski_2019}, which consumes multiple noisy magic states, or cultivation~\cite{gidney2024magicstatecultivationgrowing}, which improves one encoded state through physical operations and post-selection.

\subsubsection{Detector Construction and Decoding Confidence}
\label{sec:background-and-motivation:decoding-confidence}
QEC decoders consume protocol-defined detectors \(D_j=\bigoplus_{(q,t)\in\mathcal R_j}m_{q,t}\), whose records may span different qubits and times~\cite{Gidney_2021,McEwen_2023}. In memory circuits, \(D_j\) often compares consecutive measurements of one stabilizer. Across general FTQC protocols, it instead follows the stabilizer flow. For example, transversal CNOT maps \(S_X^{(c)}\) to \(S_X^{(c)}S_X^{(t)}\), so \(D_X=m_{X,c}^{\mathrm{pre}}\oplus m_{X,c}^{\mathrm{post}}\oplus m_{X,t}^{\mathrm{post}}\) spans both code blocks. Cultivation likewise relates root and partner-ancilla measurements during fold–unfold checks~\cite{gidney2024magicstatecultivationgrowing}. Detector construction therefore evaluates protocol-specific XORs over spatiotemporal measurement records. A detector error model (DEM) represents decoding with detector parity-check matrix \(H\), logical-observable matrix \(L\), and fault weights \(W\). Fault configuration \(e\) produces syndrome \(s=He\), logical effect \(Le\), and weight \(W^\top e\). A DEM is matchable when each fault affects at most two detectors, allowing graph-based decoding~\cite{Higgott2025sparseblossom,FusionBlossom}. The complementary gap estimates the decoding confidence by comparing minimum-weight configurations \(E_0(s)\) and \(E_1(s)\) in logical classes 0 and 1~\cite{gidney2023yokedsurfacecodes}. With \(w_i(s)=W^\top E_i(s)\),
\begin{equation}
    g_c(s) =\left|w_0(s)-w_1(s)\right|
\end{equation}
Similar weights indicate low decoding confidence, whereas a larger gap indicates higher confidence.

\begin{figure}[t]
    \centering
    \includegraphics[width=\linewidth]{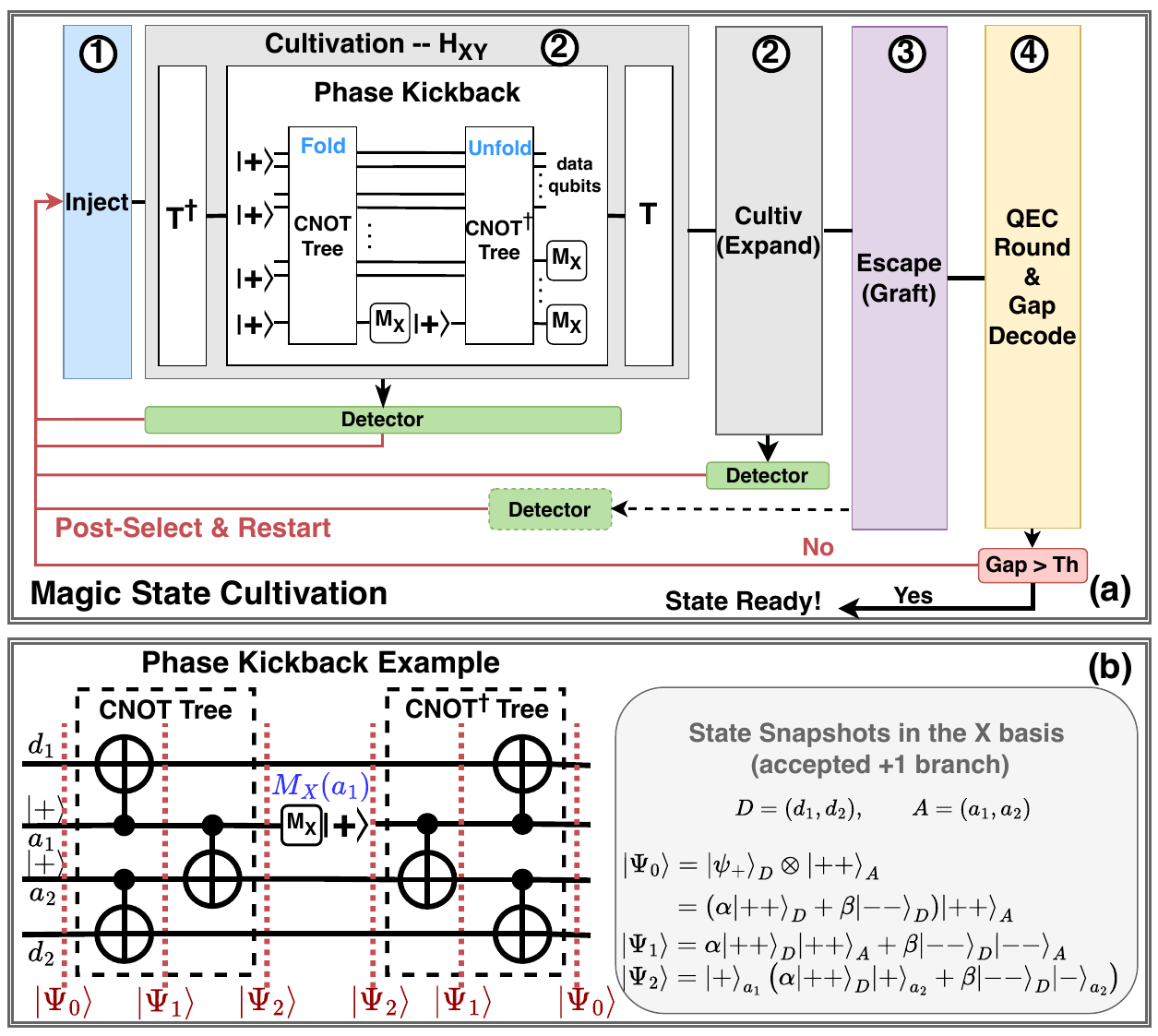}
    \caption{\textbf{Overview of magic-state cultivation.} (a) Cultivation workflow, where red paths indicate post-selection and restart. (b) Two-data-qubit phase-kickback example. Red dashed lines mark state snapshots for the accepted outcome \(M_X(a_1)=+1\), with \(D=(d_1,d_2)\) and \(A=(a_1,a_2)\).
}
    \label{fig:cultivation protocol}
\end{figure}
% \vspace{-8pt}

\subsubsection{Magic State Cultivation (MSC)}
As illustrated in Fig.~\ref{fig:cultivation protocol}(a), 
MSC~\cite{gidney2024magicstatecultivationgrowing} is a detector-triggered post-selected protocol that improves a single encoded magic state in place through three stages: Injection, Cultivation, and Escape. 
% Detector-triggered retries make its preparation latency inherently stochastic.

\paragraph{Injection.} 
The injection stage (\textcircled{1} of Fig.~\ref{fig:cultivation protocol}(a))  initializes a noisy logical magic state $\ket{T}_L$ in a small code, such as a $d=3$ color code~\cite{gidney2024magicstatecultivationgrowing}. Alternative injection circuits produce different measurement and detector structures~\cite{Chamberland_2020}. Gidney et al.\ ~\cite{gidney2024magicstatecultivationgrowing} use unitary injection, preparing the code stabilizers and logical observables while applying a physical $T$ gate mid-circuit to a qubit protected by a $Z$ stabilizer, making $X$ and $Y$ faults on the gate detectable. Any detector firing discards and restarts the attempt. Although encoded in a distance-3 code, the state has fault distance one because some single circuit faults can cause an undetectable logical error.

\paragraph{Cultivation.} Starting from the fault-distance-one injected state, the cultivation stage uses post-selected logical checks to suppress undetected logical faults to reach target fault distance \(d_{\mathrm{cultiv}}\) (the color-code distance at the end of cultivation). The target state \(\lvert T\rangle_L\) is the \(+1\) eigenstate of \(H_{XY,L}=(X_L+Y_L)/\sqrt{2}\). Cultivation therefore performs the projective measurement \(P_s^{H_{XY,L}}=(I+sH_{XY,L})/2\), where \(s\in\{+1,-1\}\), and retains only the \(+1\) outcome to project the noisy logical state onto the target eigenstate. This logical measurement can be implemented physically because the color code supports transversal \(H_{XY}\), such that \(H_{XY}^{\otimes n}\) realizes \(H_{XY,L}\) within the code space. However, directly measuring this collective non-Pauli observable is more costly than measuring a Pauli parity. Using \(H_{XY}=TXT^\dagger\), its projective measurement can instead be reduced to a collective \(X\)-parity measurement:
\begin{equation}
T^{\otimes n} P_s^X T^{\dagger\otimes n}
=
T^{\otimes n}
\frac{I + sX^{\otimes n}}{2}
T^{\dagger\otimes n}
=
\frac{I + sH_{XY}^{\otimes n}}{2}
=
P_s^{H_{XY}^{\otimes n}}.
\end{equation}
The \(H_{XY}^{\otimes n}\) measurement is then implemented by applying \(T^\dagger\) to all data qubits, nondestructively measuring their collective \(X\) parity, and applying \(T\) to restore the basis. The remaining task is to measure \(X^{\otimes n}\) without collapsing the data qubits. A standard construction uses the GHZ ancilla \(\lvert\mathrm{GHZ}_n\rangle_A=(\lvert0^n\rangle_A+\lvert1^n\rangle_A)/\sqrt{2}\). For a data state satisfying \(X_D^{\otimes n}\lvert\phi_s\rangle_D=s\lvert\phi_s\rangle_D\), transversal ancilla-to-data CNOTs produce
\begin{equation}
\ket{\mathrm{GHZ}_n}_A \ket{\phi_s}_D
\xrightarrow{\prod_i \mathrm{CNOT}_{a_i \rightarrow d_i}}
\frac{\ket{0^n}_A + s\ket{1^n}_A}{\sqrt{2}}
\ket{\phi_s}_D .
\end{equation}
The eigenvalue \(s\) is thus kicked back onto the relative phase of the GHZ state. The \(X\)-basis parity of the ancilla measurements reveals \(s\) while leaving the data state unchanged. However, explicitly preparing and verifying a large GHZ state is costly, motivating the fold-based implementation~\cite{rosenfeld2025magicstatecultivationsuperconducting}. In this implementation (\textcircled{2} of Fig.~\ref{fig:cultivation protocol}(a)), each data qubit is paired with an adjacent partner drawn from the measurement-ancilla sites already present in the color-code layout. Reusing these sites preserves the patch footprint and local connectivity while allowing the same ancillas to transition back to the surrounding QEC rounds. Each partner-to-data CNOT obeys \(\mathrm{CNOT}_{a_i\rightarrow d_i}^{\dagger}X_{a_i}\mathrm{CNOT}_{a_i\rightarrow d_i}=X_{a_i}X_{d_i}\), coherently encoding the data-qubit \(X\) information in its partner. The partners are then connected by a spanning CNOT tree. Since each parent-to-child CNOT maps \(X_p\) backward to \(X_pX_c\), the tree maps the root measurement to \(\prod_iX_{a_i}\), and the partner CNOTs extend it to \((\prod_iX_{a_i})(\prod_iX_{d_i})\). As every partner starts with \(X=+1\), measuring the root therefore measures \(\prod_iX_{d_i}\). In Fig.~\ref{fig:cultivation protocol}(b), this reduces to the desired \(X_{d_1}X_{d_2}\) parity. The root measurement provides the first logical-check outcome. Reinitializing the root in \(\lvert+\rangle\) and reversing the CNOT sequence restores the data state and disentangles the partners, whose final \(X\)-basis measurements embed a time-reversed second check and additional flags. Detector construction combines these outcomes into deterministic parities, and any fired detector rejects the attempt, so post-selection operates on detectors rather than raw measurements. To further increase the fault distance, cultivation can expand the color code by initializing added data qubits as Bell pairs, followed by superdense stabilizer cycles~\cite{Lacroix_2025} that initialize new stabilizers and limit correlated faults between logical checks. Further expansion can target larger \(d_{\mathrm{cultiv}}\), but rapidly increases the post-selection cost, making \(d_{\mathrm{cultiv}}\in\{3,5\}\) the practical choices.
\paragraph{Escape}
The escape stage transfers the cultivated state from the small color code into a much larger code suitable for error correction without full post-selection. The resulting distance \(d_{\mathrm{escape}}\) does not remove logical faults missed during cultivation, but suppresses faults introduced during escape and subsequent storage. During grafting, shots firing selected detectors in the residual color-code region (the dashed block in Fig.~\ref{fig:cultivation protocol}(a)) are discarded as the region is removed or decomposed, making the DEM matchable. The complementary gap is then computed after decoding and compared with a tunable threshold, trading lower output LER for higher retry overhead. Supporting this calculation requires representing the logical observable as an additional detector connected to logical-flipping errors. The logical detector can have high degree, increasing decoding latency and hardware cost. While decoding proceeds, the patch remains protected through syndrome-extraction rounds.
\paragraph{Variants and Use Cases}
\label{sec:background:msc:usecase}
Magic-state cultivation is not tied to one code or circuit. It mainly requires efficiently and fault-tolerantly measuring a logical operator whose eigenstate is the target magic state. The original color-code construction performs this check transversally. Later variants alter the code, connectivity, or escape procedure~\cite{9kys-3whh,hirano2026efficientmagicstatecultivation,sahay2026foldtransversalsurfacecodecultivation}. Cultivated states can be consumed directly or feed subsequent distillation or synthillation~\cite{Campbell_2017}. Direct use avoids another factory but requires cultivation to meet the application error budget~\cite{campbell2026resourceestimationefficientcompilation,gidney2025factor2048bitrsa}, typically through stricter post-selection and more retries. As an upstream source, cultivation tolerates a higher LER, improving acceptance and throughput at the cost of downstream operations and factory resources~\cite{xu2026distilling,cain2026shorsalgorithmpossible10000}.

\begin{figure}[htbp]
    \centering
    \includegraphics[width=\linewidth]{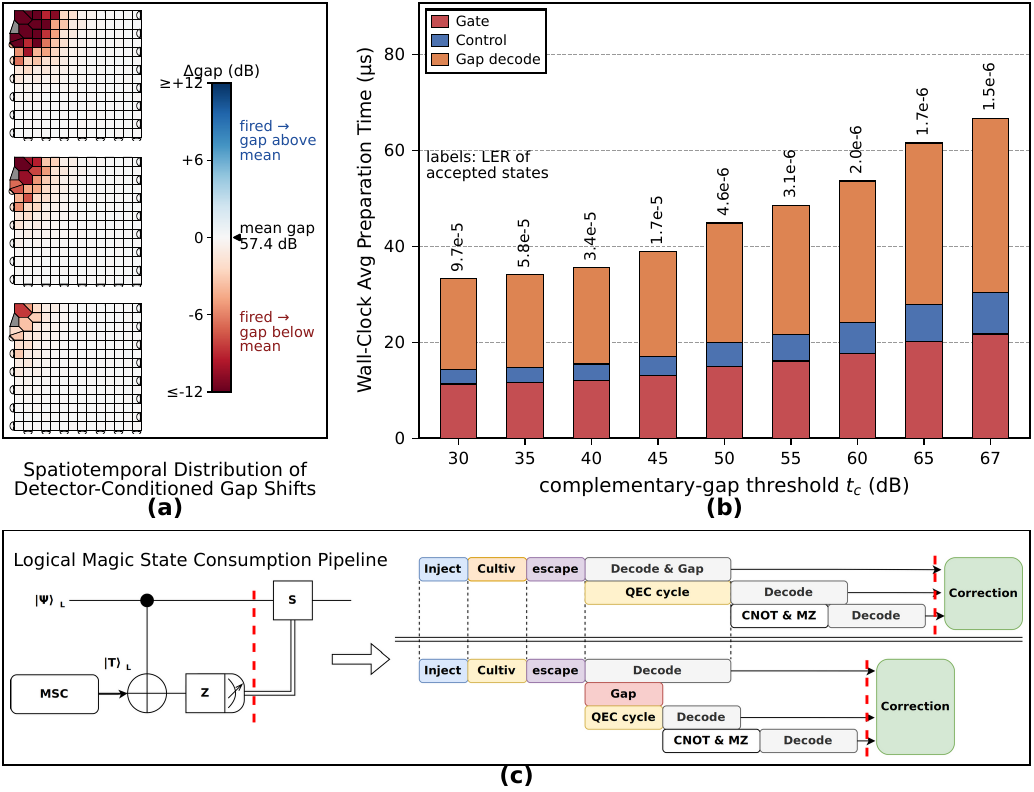}
    \caption{
            (a) Spatiotemporal detector-conditioned gap shift,
                $\Delta g(d)=\mathbb{E}[g_c\mid s_d=1]-\mathbb{E}[g_c]$, where $s_d=1$
                denotes detector firing.
            (b) Wall-clock average preparation-latency (including accumulated latencies for all retries until first success) breakdown across complementary-gap
                thresholds $t_c$ for $d_{\mathrm{cultiv}}=3$,
                $d_{\mathrm{escape}}=15$, and $p=10^{-3}$; labels report the
                accepted-state LER; gate time from~\cite{2023}; control latency assumed to be $3.1\mu s$~\cite{Caune2026RealtimeQEC}.
            (c) Logical magic-state consumption pipeline, showing how an earlier gap
                decision overlaps downstream operations with the remaining decoding.
            }
    \label{fig:motivation}
\end{figure}
\subsection{Motivation}
\label{sec:background-motivation-and-formalism:motivation}
Existing quantum-control systems support measurement-conditioned feedback~\cite{xu2023qubic20extensibleopensource,10.1145/3725843.3756048}, while recent QEC systems integrate real-time decoding~\cite{Caune2026RealtimeQEC,liu2026scalableopensourceqecsubmicrosecond}, yet neither provides a protocol-aware runtime for end-to-end execution of detector-driven FTQC state preparation. We therefore (i) bridge this execution gap with a configurable hardware runtime instantiated for magic-state cultivation, and (ii) reduce cultivation decision latency through partial-detector early acceptance.
% \subsubsection{\textbf{Architectural Challenge: From FTQC Abstractions to Executable Protocols}}
% FTQC architecture and compilation frameworks~\cite{Watkins2024highperformance,pflieger2026harvestresourceawarequantumcompilation,kan2025sparosurfacecodepaulibasedarchitectural} describe complete computations through coarse-grained logical operations, while physical-control systems~\cite{10.1145/3725843.3756048} increasingly adopt distributed tree-grid organizations for measurement and feedback, and recent QEC systems further integrate real-time decoding~\cite{caune2024demonstratingrealtimelowlatencyquantum,liu2026scalableopensourceqecsubmicrosecond}. Progress at both ends of the stack does not by itself bridge the gap between them: a logical operation is still not automatically lowered into the online control required by its underlying protocol. As FTQC protocols mature, their time-critical semantics need to be progressively realized in a hardware runtime between logical software and physical control to improve efficiency and meet latency requirements. For example, a scheduler may issue \textsc{PrepareMagicState} as one operation, while the system must construct detectors, evaluate protocol conditions, synchronize controllers, and manage retries efficiently before returning a usable state.
\subsubsection{Architectural Challenge: From FTQC Abstractions to Executable Protocols.}
FTQC architecture and compilation frameworks~\cite{Watkins2024highperformance,pflieger2026harvestresourceawarequantumcompilation,kan2025sparosurfacecodepaulibasedarchitectural} express
computations through coarse-grained logical operations, while physical-control
and QEC systems provide measurement feedback and real-time decoding.
However, these operations are not automatically lowered into protocol-level
control. As FTQC protocols mature, their time-critical semantics require a
runtime between logical software and physical control. For example, a scheduler
may issue \textsc{PrepareMagicState} as one operation, while the system must
construct detectors, evaluate conditions, synchronize controllers, and manage
retries before returning the state.

\noindent\textbf{Insight and Our Approach.}
% We separate protocol-specific semantics from latency-critical execution.
% Software offline compiles protocol's static rules into per-controller
% instructions. A configurable hardware runtime executes these
% instructions online to construct detectors, evaluate protocol decisions,
% coordinate controllers, and interface with the decoder and physical
% control cores. This keeps the measurement--decision loop (detector instead of raw measurement triggered) in hardware,
% while supporting different FTQC protocols, and control backends
% without hard-coding everytime. We take FTQC state preparation as an initial instance of this broader
% runtime layer, with cultivation as its first protocol. Compared with
% multi-patch logical protocols such as distillation, cultivation is
% self-contained enough to isolate the runtime problem, yet exercises
% nontrivial detector construction, online postselection, decoding-based
% acceptance, and distributed synchronization. We therefore develop
% MagicFirm, a portable hardware runtime that integrates with existing
% physical-control systems to execute cultivation online from offline-compiled
% instructions.
Inspired by microprogrammed control in classical processors, we separate
static protocol semantics from latency-critical execution. Software compiles
protocol rules into per-controller microinstructions offline, while a
configurable hardware runtime executes them online to coordinate detector
construction, protocol decisions, decoding, and physical control. This keeps
the detector-driven control loop in hardware without hard-coding a specific
protocol or control backend. We instantiate this runtime for cultivation,
which exercises post-selection, decoding-based acceptance, and distributed
retry control.

\subsubsection{\textbf{Performance Challenge: Cultivation Decision Latency}}
\label{sec:motivation:performance challenge}
% With this execution layer in place, cultivation performance must be
% evaluated over the complete quantum--classical path. Recent resource
% estimates and schedulers model cultivation through qubit-round or
% logical-cycle costs and stochastic magic-state availability, rather than
% resolving the detector processing, decoding, complementary-gap
% evaluation, and feedback latency within each
% attempt~\cite{hofmeyr2026puremagicdynamicschedulerlattice,pflieger2026harvestresourceawarequantumcompilation,campbell2026resourceestimationefficientcompilation}. In a real system, an accepted state cannot be
% released until this decision path completes. We therefore build a
% cycle-accurate cultivation model combining circuit execution with
% component-level control and decoding latencies. As shown in
% \figref{fig:motivation}(b), classical decision processing is
% non-negligible under the evaluated configuration. Although its relative
% contribution varies across configurations, reducing it can make accepted
% states available earlier to downstream computation.
With this execution layer in place, cultivation performance must be
evaluated over the complete quantum--classical path. Recent resource
estimates and schedulers model cultivation through qubit-round or
logical-cycle costs and stochastic magic-state availability~\cite{hofmeyr2026puremagicdynamicschedulerlattice,
pflieger2026harvestresourceawarequantumcompilation,
campbell2026resourceestimationefficientcompilation}, without
resolving detector processing, decoding, complementary-gap evaluation,
and feedback within each
attempt.
In a real system, an accepted state cannot be released until this decision
path completes, and its latency accumulates across retries. We therefore
build a cycle-accurate cultivation model combining circuit execution with
component-level control and decoding latencies. In
\figref{fig:motivation}(b), increasing $t_c$ from 30 to 67\,dB lowers the
accepted-state LER from $9.7\times10^{-5}$ to $1.5\times10^{-6}$, but
raises average preparation time from approximately 31.5 to 66.7\,$\mu$s.
Across this sweep, classical decision processing contributes roughly two thirds of this
latency, making it an important target for earlier state availability.

\noindent\textbf{Insight and Our Approach.}
% Cultivation grows a small code into a larger escape code, yet the initial
% code determines the protocol's fault distance. Thus, detectors need not
% contribute equally to the final complementary gap. Statistical profiling confirms
% that gap information is concentrated in a relatively small detector
% subset, as shown in \figref{fig:motivation}(a). Its resulting smaller DEM can be
% decoded faster, providing an earlier estimate of the final gap. We exploit this observation with an offline-constructed partial detector mask and two-stage early acceptance. At runtime, attempts with a sufficiently high
% partial gap are accepted early, while the remainder use complete-gap
% evaluation. This separates state availability from complete decoding,
% enabling overlap with downstream processing, as illustrated in
% \figref{fig:motivation}(c), without changing the quantum protocol or the underlying
% decoder.
Cultivation grows a fault-distance-setting code into a larger escape code,
so detectors across regions and rounds need not contribute equally to the
final complementary gap. We test this with a diagnostic measuring the gap
shift conditioned on each detector firing. In \figref{fig:motivation}(a),
only 48 of 1,027 detectors (4.7\%) reduce the conditional mean gap by at
least 6\,dB, while the top 10\% account for 83.2\% of the summed negative
shifts. This diagnostic characterizes the opportunity rather than selecting
the mask; our offline algorithm instead profiles logical ambiguity and
preserves its surrounding error structure. To translate this reduction into
system-level latency savings, the partial and complete decoders run in
parallel at runtime. A sufficiently high partial gap allows downstream
execution to begin early, while other attempts await the complete-gap
decision. The complete decoder still finishes to supply the correction, but
after early acceptance its remaining latency overlaps downstream operations
and is hidden from the state-availability critical path, as illustrated in
\figref{fig:motivation}(c). This reduces decision latency without changing
the quantum protocol or decoder.

\section{MagiCFirm System}
\label{sec:magic firm sys}

\subsection{Control System Design}
\label{sec:magic firm sys:controlsys}
\begin{figure*}
    \centering
    \includegraphics[width=515.52pt]{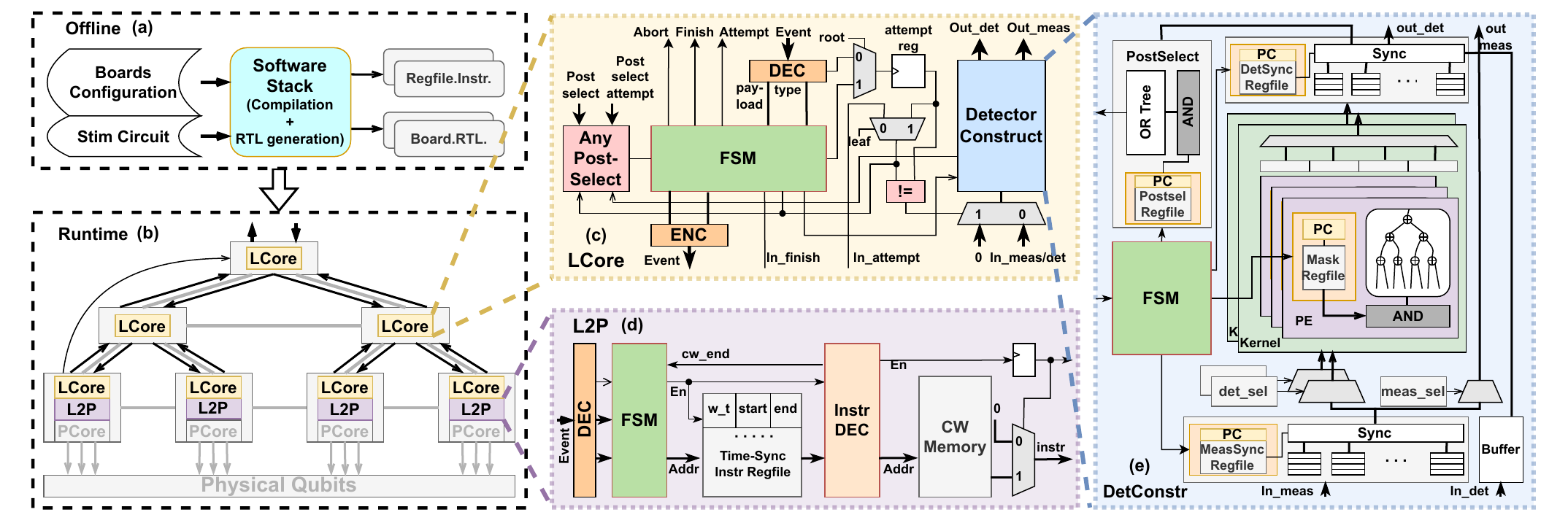} % 0.95\linewidth]
    \caption{\textbf{MagiCFirm control-system architecture.} The left panels show offline compilation and hierarchical runtime deployment over an existing physical-control substrate, where PCore denotes the physical control core that directly controls the underlying qubits; the center panels detail the LCore and L2P architectures; and the right panel expands the configurable detector-construction datapath. Gray components denote the existing physical-control system, while colored blocks and black interconnects denote MagiCFirm extensions.}
    \label{fig:control system arch}
    \Description{}
\end{figure*}
As illustrated in Fig.~\ref{fig:control system arch}(b), we assume an underlying control system that provides synchronized clocks and coordinated physical operations across control boards, as demonstrated by existing distributed control architectures~\cite{10.1145/3725843.3756048}. 
MagiCFirm is designed as an FTQC control layer that integrates with existing physical quantum-control systems. 
On top of our proposed system, MagiCFirm introduces a hierarchy of logical control cores (LCores) for the real-time construction of configurable non-local and cross-round detectors, post-selection, and QEC-level protocol control. As illustrated in Fig.~\ref{fig:control system arch}(a), an offline compiler maps the quantum circuit and system configuration onto per-board microprogrammed-instructions and configuration, allowing the same parameterized LCore architecture to support both monolithic and distributed deployments. While this work focuses on magic-state cultivation, the architecture is not cultivation-specific: it targets protocols whose quantum operations can be compiled offline while runtime execution is coordinated through QEC-level events, including detector-driven post-selection and retry. A generic interface to the underlying physical control cores (PCores) further decouples the LCores from platform-specific control implementations.

\subsubsection{Event-Based Protocol Control} 
Post-selective QEC protocols require runtime decisions to trigger synchronized execution changes across multiple control boards. This is particularly strict in magic-state cultivation: as the code expands from a smaller cultivation patch to a larger code, previously idle qubits must begin operations at the prescribed circuit boundary. Such transitions can be triggered by post-selection outcomes generated at individual LCores. To propagate these local decisions with low latency, designated stage LCores send their outcomes directly to the root through a low-bandwidth fast path (Fig.~\ref{fig:control system arch}(b)), bypassing the detector hierarchy. The root converts the outcome into a global control event and broadcasts it down the balanced LCore tree. Together with the globally synchronized clock of the physical-control substrate, this tree provides deterministic event-arrival latency across the hierarchy, enabling synchronized protocol transitions without peer-to-peer coordination.

% As shown in Fig~\ref{fig:control system arch}(c), to encode both the requested transition and its execution instance, each event carries an \textit{event type} and an \textit{attempt tag}. The current implementation defines three event types: \texttt{START} initiates execution, \texttt{ABORT} rejects the current attempt and starts a retry, and \texttt{FINISH} terminates execution and initiates its drain. Upon receiving an event, each LCore updates its local FSM and redirects the instruction pointers of protocol-dependent components, including detector-construction regfiles; leaf LCores additionally redirect physical-control execution to the corresponding precompiled sequence. During an \texttt{ABORT}, measurement and detector data from the old attempt may still propagate upward while the new attempt propagates downward. Each LCore therefore accepts only data matching its local attempt tag, discarding stale results as the new attempt reaches that board. This allows a new attempt to start without waiting for all data from the previous attempt to leave the hierarchy. For \texttt{FINISH}, each LCore returns to idle once its detector datapath reports that the subtree has drained. Beyond these operations, additional event types can redirect LCores to other precompiled microprogram entries, enabling runtime mode switches such as from waiting QEC to a consumption sequence without host intervention or runtime instruction generation.
As shown in Fig.~\ref{fig:control system arch}(c), each event carries an
\textit{event type} and an \textit{attempt tag} to identify the requested
transition and execution instance. We define three event types:
\texttt{START} begins execution, \texttt{ABORT} rejects the current attempt
and starts a retry, and \texttt{FINISH} terminates execution and initiates
draining. On receipt, each LCore updates its local FSM and redirects the
instruction pointers of protocol-dependent components, including
detector-construction regfiles; leaf LCores also redirect physical control to
the corresponding precompiled sequence. During an \texttt{ABORT}, old
measurement and detector data may still propagate upward as the new attempt
propagates downward. Each LCore therefore accepts only data matching its local
attempt tag and discards stale results when the new attempt arrives. This
starts a retry without waiting for old data to drain from the hierarchy. For
\texttt{FINISH}, each LCore returns to idle once its detector datapath reports
that the subtree has drained. Additional event types can redirect LCores to
other precompiled microprogram entries, switching operational modes (e.g.,
from idle QEC to a consumption sequence) without host intervention or
runtime instruction generation.

\subsubsection{Logical-to-Physical Control Interface}
% MagiCFirm decouples LCore execution from physical control through a two-level Logical-to-physical instruction mapping. As shown in Fig.~\ref{fig:control system arch}(d), each leaf stores backend-specific physical-control sequences in a command-word (CW) memory. A CW may encode a native PCore instruction, waveform data, or other control operation without exposing its semantics to the LCore. A time-synchronized instruction regfile defines the virtual instruction space, where each entry $(w_t,\mathrm{start},\mathrm{end})$ specifies the cycles to wait before streaming CWs from $\mathrm{start}$ through $\mathrm{end}$ to the PCore. Runtime events redirect the virtual instruction pointer, with each entry controlling the timing and addressing of the CW memory. This mapping separates protocol-level control from backend-specific command representation, allowing backend adaptation without modifying the LCore architecture.
MagiCFirm decouples LCore execution from physical control through a two-level
logical-to-physical instruction mapping. As shown in
Fig.~\ref{fig:control system arch}(d), each leaf stores backend-specific
physical-control sequences in command-word (CW) memory. A CW may encode a
native PCore instruction, waveform data, or another control operation without
exposing its semantics to the LCore. A time-synchronized instruction regfile
defines the logical instruction space; each entry
$(w_t,\mathrm{start},\mathrm{end})$ waits $w_t$ cycles before streaming the
specified CW range to the PCore. Runtime events redirect the logical
instruction pointer to select the appropriate entry. This mapping separates
protocol-level control from backend-specific commands, enabling backend adaptation without modifying the LCore.

\subsubsection{Configurable Detector Construction}
% MagiC-Firm constructs each detector at the lowest LCore with access to all required measurements, avoiding centralized raw-measurement collection at the root. Each LCore forwards constructed detectors and only residual measurements required at higher levels. InFig.~\ref{fig:control system arch}(e), incoming measurements are FIFO-buffered and released by measurement time index using compiler-generated synchronization masks. The datapath exploits the spatial sparsity and temporal locality of detector dependencies: each of the $K$ kernels represents a detector channel (spatial index) and selects its required measurements, while parallel PEs use configurable masks to XOR them into the partial parities of different relative-time detector instances. These parities are maintained in circular registers and reused as detectors complete. Although dependencies may span non-adjacent qubits or control boards, syndrome-extraction circuits typically restrict them to sparse measurement sets across few rounds, bounding kernel fan-in and temporal PE count. Constructed detectors are then synchronized by detector time index before upward propagation, while designated detectors directly trigger the post-selection fast path to the root.
MagiCFirm constructs each detector at the lowest LCore containing all required
measurements, avoiding raw-measurement collection at the root. Each LCore
forwards completed detectors and only residual measurements needed at higher
levels. Fig.~\ref{fig:control system arch}(e) shows incoming measurements
FIFO-buffered and released by time index using compiler-generated
synchronization masks. The datapath exploits spatial sparsity and temporal
locality: each of the $K$ kernels represents a detector channel and selects
its required measurements, while parallel PEs apply configurable XOR masks to
form partial parities for different relative-time instances. These parities
are maintained in circular registers and reused as detectors complete.
Although dependencies may span non-adjacent qubits or control boards,
syndrome-extraction circuits typically restrict them to sparse measurement
sets over few rounds, bounding kernel fan-in and temporal PE count. Completed
detectors are synchronized by time index for upward propagation, while
designated detectors directly trigger the post-selection fast path to the
root.

\subsection{Partial Gap Estimation}
\label{sec:magic firm sys:partialgap}
% This section proposes partial gap estimation to reduce the classical processing cost of complementary-gap estimation. Offline, profiling shots collected from simulation or hardware are decoded to derive the statistics used for \textbf{partial mask construction}, after which \textbf{DEM contraction} produces a reduced decoding problem. At runtime, the partial decoder decodes only the masked syndrome to produce the partial gap.
This part proposes partial gap estimation to reduce the classical processing cost of complementary-gap estimation. Offline, profiling shots collected from simulation or hardware are decoded to derive the statistics used for \textbf{partial mask construction}, after which \textbf{DEM contraction} produces a reduced decoding problem. At runtime, it decodes only the masked syndrome to obtain the partial gap.

\subsubsection{Partial Mask Construction}
\label{sec:magic firm sys:partialgap:partialmask}

\begin{algorithm}[t]
\caption{Partial-Mask Construction}
\label{alg:partial-mask-construction}

\SetKwInOut{Input}{Input}
\SetKwInOut{Output}{Output}
\SetKwFunction{TopK}{TopK}

\Input{
$\mathcal{D}$: detector error model; \\
$\mathcal{S}$: profiling shots; \\
$K_Q$: base-mask budget
}
\Output{Partial detector mask $M$}

\BlankLine
% \textit{\textbf{\textsc{Logical Ambiguity Profile}}}\;
% \textit{\# \textsc{Logical Ambiguity Profile}}\:
\textbf{Phase I: Logical Ambiguity Profile}\;

\ForEach{$s \in \mathcal{S}$}{
    $\bigl(E_0(s),w_0(s)\bigr),
     \bigl(E_1(s),w_1(s)\bigr)
     \leftarrow
     \textsc{Decode}(\mathcal{D},s)$\;

    $Z_s \leftarrow E_0(s)\oplus E_1(s)$\;
    \nllabel{line:logical-cycle}

    $r_s \leftarrow
    \exp\!\left(-|w_0(s)-w_1(s)|\right)$\;
    \nllabel{line:ambiguity-weight}
}

\ForEach{$d \in \operatorname{Det}(\mathcal{D})$}{
    $Q(d) \leftarrow
    \displaystyle
    \frac{
        \sum_{s\in\mathcal{S}}
        r_s\,\mathbf{1}
        [d\in\operatorname{supp}_d(Z_s)]
    }{
        \sum_{s\in\mathcal{S}} r_s
    }$\;\nllabel{line:lap-score}
}

$M \leftarrow \TopK(Q,K_Q)$\;
\nllabel{line:lap-topk}

\BlankLine
% \textbf{\textsc{Structural Closure}}\;
% \textit{\# \textsc{Structural Closure}}\:
\textbf{Phase II: Structural Closure}\;

\Repeat{$A=\varnothing$}{
    $F \leftarrow
    \left\{
        d\notin M :
        \exists e\in\mathcal{E}_{\mathrm{cross}}(M),
        \ d\in e
    \right\}$\;
    \nllabel{line:closure-frontier}

    $A \leftarrow
    \left\{
        d\in F :
        S(M\cup\{d\}) < S(M)
    \right\}$\;

    $M \leftarrow M\cup A$\;
}

\Return{$M$}\;

\end{algorithm}

Motivated by the non-uniform distribution of gap-relevant information identified in Sec.~\ref{sec:motivation:performance challenge}, the purpose is to construct a compact detector mask for gap estimation. This introduces two challenges. First, \textit{information relevance}: under a limited detector budget, the mask should retain detectors informative of competing logical hypotheses rather than simply preserving arbitrary syndrome information. Second, \textit{structural completeness}: masking out part of an error component can remove the context needed to interpret the retained detectors. Algorithm~\ref{alg:partial-mask-construction} addresses these challenges in two phases. \textbf{\textit{Logical Ambiguity Profile} (LAP)}  identifies detectors associated with competing logical explanations to construct an information-rich base mask, while \textbf{\textit{Structural Closure}} augments this mask to reduce error components split across the mask boundary.

\paragraph{Logical Ambiguity Profile}
\label{sec:magic firm sys:partialgap:partialmask:LAP}
% To identify detectors informative of logical ambiguity, we profile the two
% error configurations defining the complete complementary gap. For each shot
% $s$, complete decoding yields the minimum-weight configurations $E_0(s)$
% and $E_1(s)$ from the two logical classes. While their weight separation
% $g_c(s)$ quantifies logical ambiguity, it does not reveal where this
% ambiguity is represented in detector space. We therefore take their
% symmetric difference (Alg.~\ref{alg:partial-mask-construction},
% Line~\ref{line:logical-cycle}),
% % \begin{equation}
% %     Z_s = E_0(s)\oplus E_1(s)
% % \end{equation}
% which cancels error components shared by the two configurations and retains those on which they differ. Since they produce the same syndrome with different logical effect, $Z_s$ is a non-trivial logical cycle separating the two decoding hypotheses. We then define its detector support as
For each profiling shot $s$, let $E_0(s)$ and $E_1(s)$ be the competing
configurations defined in Sec.~\ref{sec:background-and-motivation:decoding-confidence}. Their separation $g_c(s)$
quantifies logical ambiguity but does not locate it in detector space. We
therefore form their symmetric difference
$Z_s=E_0(s)\oplus E_1(s)$
(Alg.~\ref{alg:partial-mask-construction},
Line~\ref{line:logical-cycle}), which removes shared error components.
Because $E_0(s)$ and $E_1(s)$ produce the same syndrome but different logical
effects, $Z_s$ is a non-trivial logical cycle whose detector support localizes
the ambiguity. We define this support as

\begin{equation}
    \operatorname{supp}_{d}(Z_s) \triangleq \left\{d \mid \exists\, e \in Z_s,\ d \in e \right\}
    = \bigcup_{e \in Z_s} e
\end{equation}
which contains the detectors incident to at least one error component in the
logical cycle. Not all cycles are equally informative: those arising from
nearly degenerate configurations indicate greater logical ambiguity. Assuming
$w_0(s)\geq w_1(s)$ without loss of generality, negative log-likelihood
weights give
\begin{equation}
    r_s
    \triangleq
    \frac{P(E_0(s)\mid s)}{P(E_1(s)\mid s)}
    =
    \exp[-(w_0(s)-w_1(s))]
    =
    \exp[-g_c(s)] .
\end{equation}
We use $r_s$ as the ambiguity weight
(Alg.~\ref{alg:partial-mask-construction},
Line~\ref{line:ambiguity-weight}). 
Near-degenerate configurations receive
weights close to one, while well-separated configurations are down-weighted.
The \textsc{LAP} score $Q(d)$
(Algorithm~\ref{alg:partial-mask-construction}, Line~\ref{line:lap-score})
measures the ambiguity-weighted frequency with which detector $d$ participates
in these cycles. We initialize $M$ with the $K_Q$ detectors having the highest
scores (Line~\ref{line:lap-topk}).

\begin{figure}[htbp]
    \centering
    \includegraphics[width=0.6\linewidth]{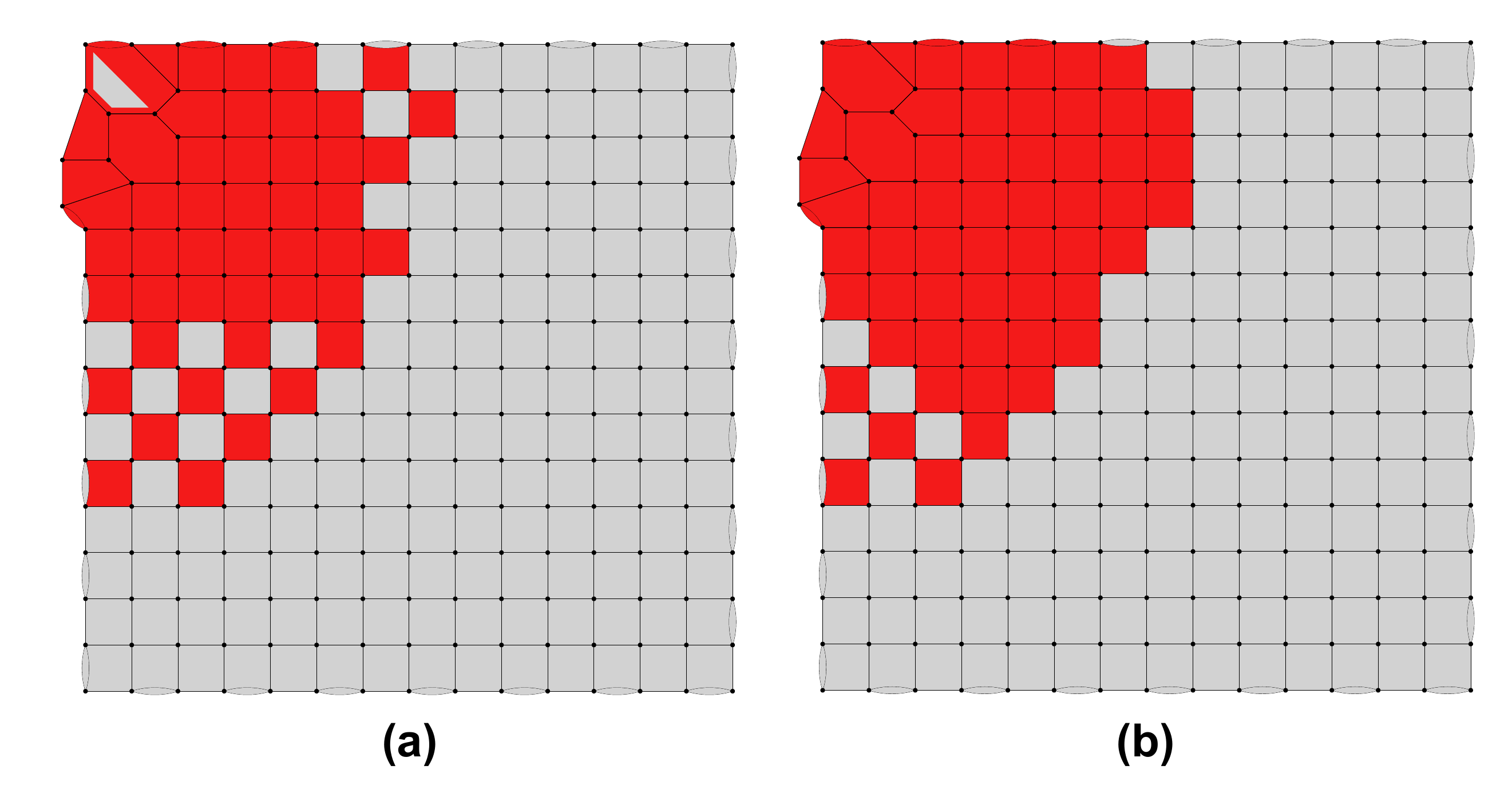}
    \caption{\textbf{Partial mask geometric visualization.}
    A representative time slice of the partial mask for
    $d_{\mathrm{cultiv}}=3$, $d_{\mathrm{escape}}=15$, $p=10^{-3}$,
    and $K_Q=200$.
    Red and gray cells denote retained and excluded detectors.
    (a) The LAP-selected base mask.
    (b) Structural closure augmented mask.}
    \label{fig:LAP_with_without_closure_mask}
\end{figure}

\paragraph{Structural Closure}
Fig.~\ref{fig:LAP_with_without_closure_mask}(a) shows a representative LAP
base mask. Although LAP concentrates detectors in the gap-relevant region,
detector-wise selection leaves interior holes and cuts error components across
the mask boundary. The resulting partial observations may force the decoder
toward more complex explanations, introducing ambiguity absent from the
complete detector set. We capture this missing structural context using \emph{crossing errors}. An error component $e$ is crossing if
$e\cap M\neq\varnothing$ and $e\cap\overline{M}\neq\varnothing$;
let $\mathcal{E}_{\mathrm{cross}}(M)$ denote their set. We quantify
the structural incompleteness of mask $M$ by the probability-weighted crossing
mass
\begin{equation}
    S(M)=\sum_{e\in\mathcal{E}_{\mathrm{cross}}(M)}p_e,
\end{equation}
where $p_e$ is the probability of error component $e$. Starting from the LAP
mask, structural closure iteratively considers excluded detectors incident to
crossing errors and adds those whose inclusion reduces $S(M)$
(Algorithm~\ref{alg:partial-mask-construction},
Line~\ref{line:closure-frontier}). Because adding a detector may remove
existing crossings while creating others, this criterion prevents
indiscriminate mask expansion. As shown in
Fig.~\ref{fig:LAP_with_without_closure_mask}(b), closure fills many interior
holes and regularizes the boundary without using explicit geometric
information.

% \begin{table}[htbp]
% \centering
% \renewcommand{\arraystretch}{1.3}
% \begin{tabularx}{\columnwidth}{|c|X|c|}
% \hline
% \textbf{Correlation Level} & \textbf{Pauli Errors} & \textbf{Factor ($\alpha$)} \\ \hline
% 0 & $IX, IZ, XI, XX, ZI, ZZ$ & 2 \\ \hline
% 1 & $IY, XZ, YI, ZX$ & 4 \\ \hline
% 2 & $XY, YX, YY, YZ, ZY$ & 7 \\ \hline
% \end{tabularx}
% \vspace{1ex}
% \caption{Classification of the $15$ non-trivial CNOT Pauli errors into cumulative correlation levels ($L_0 \subset L_1 \subset L_2$). 
%          The factor $\alpha$, which determines the STEPG node density via $k = \alpha n$, is defined recursively as $\alpha_i = \alpha_{i-1} + \lceil m/2 \rceil$, where $m$ is the number of additional correlated errors introduced at each level (4 for $L_1$; 5 for $L_2$).
%          This hierarchy applies specifically to the CNOT gate; other two-qubit gates (e.g., CZ or iSWAP) would necessitate a different hierarchical approach.}
% \label{tab:correlation-level}
% \end{table}

\subsubsection{DEM Contraction}
\label{sec:magic firm sys:partialgap:demcontract}

\begin{algorithm}[t]
\caption{Graph-Based DEM Contraction}
\label{alg:graph-contraction}

\SetKwInOut{Input}{Input}
\SetKwInOut{Output}{Output}

\Input{
$G=(V,E)$: decoding graph with edge weight $w(e)$ and logical effect $\ell(e)$; \\
$V_M\subseteq V$: retained detector vertices; \\
$\tau$: synthetic-edge weight cutoff
}
\Output{Contracted decoding graph $G_M$}

\BlankLine
\textbf{Phase I: Reduced-Fault-Set Initialization}\;
$V_{\bar M}\leftarrow V\setminus V_M$, \quad
$\tilde{E}\leftarrow\varnothing$\;

$\hat{E}\leftarrow
\{e=(u,v)\in E\mid u, v \in V_M \cup \{\partial\}\}$\;
\nllabel{line:Ehat}

\BlankLine
\textbf{Phase II: Hidden-Space Contraction}\;
\ForEach{$\hat{v}\in V_M$}{
    $W(v)\leftarrow\infty,\ \forall v\in V_{\bar M};
    \quad Q\leftarrow\textsc{PriorityQueue}$\;

    \ForEach{$e=(\hat{v},v)\in E,\ v\in V_{\bar M}$}{
        $W(v)\leftarrow w(e), \textsc{Enqueue}(Q,(w(e),v,\ell(e)))$\;
    }

    \While{$Q\neq\varnothing$}{
        $(w_v,v,\ell_v)\leftarrow\textsc{Dequeue}(Q)$\;

        \ForEach{$e=(v,u)\in E,\ u\neq\hat{v}$}{
            $w'\leftarrow w_v+w(e),\quad
            \ell'\leftarrow\ell_v\oplus\ell(e)$\;

            \uIf{$u\in V_M\setminus\{\hat{v}\}\lor u=\partial$}{
                $\tilde{E}[\hat{v},u]\leftarrow\min_w\{\tilde{E}[\hat{v},u],(w',\ell')\}$\;
            }
            \ElseIf{$w'<W(u)$}{
                $W(u)\leftarrow w', \textsc{Enqueue}(Q,(w',u,\ell'))$\;
            }
        }
    }
}

\BlankLine
\textbf{Phase III: Weight Truncation}\;
$\tilde{E}_{\tau}\leftarrow
\{e\in\tilde{E}\mid
w(e)<\tau \land w(e)<w_{\hat{E}}(e)\}$\;
\nllabel{line:truncate}

\BlankLine
\textbf{Phase IV: Feasibility Repair}\;
$\mathcal{C}\leftarrow
\textsc{ConnectedComponents}(V_M,\hat{E}\cup\tilde{E}_{\tau})$\;

\ForEach{$V_i\in\mathcal{C}$ with no path to $\partial$}{
    $e^*\leftarrow
    \arg\min_{e=(v,\partial)\in\tilde{E},\,v\in V_i}w(e)$, 
    $\tilde{E}_{\tau}\leftarrow\tilde{E}_{\tau}\cup\{e^*\}$\;
}

$E_M\leftarrow\hat{E}\cup\tilde{E}_{\tau}$\;
\Return{$G_M=(V_M,E_M)$}\;

\end{algorithm}

\begin{figure}[htbp]
    \centering
    \includegraphics[width=\linewidth]{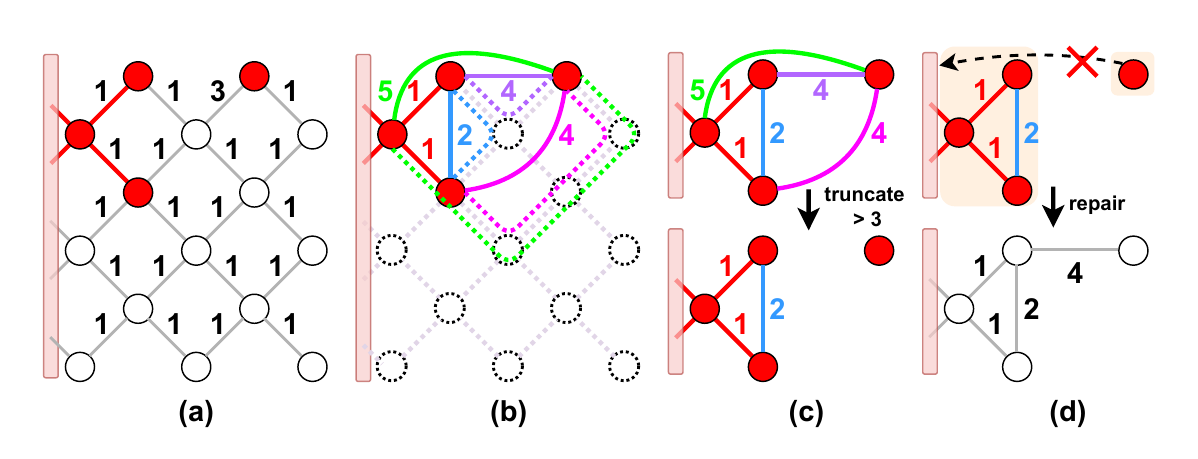}
    \caption{\textbf{Graph-based DEM contraction.}
    Red, gray vertices denote retained, masked-out detectors. The red block is the boundary. Each colored solid edge denotes an effective fault, with the same-colored dashed edges showing original hidden-space path. Panels (a)--(d) correspond to 4 phases.}
    \label{fig:graph contraction}
\end{figure}

% The computational cost of DEM-based decoding depends strongly on the size and connectivity of the underlying DEM~\cite{Higgott2025sparseblossom,wu2023fusionblossomfastmwpm,wu2025minimumweightparityfactordecoder}. Consequently, reducing the detector observations alone does not fully realize the computational benefit of partial decoding if the decoder still operates on the original DEM. Given the detector mask $M$, we therefore construct a smaller effective DEM over the kept detectors, reducing the decoding problem itself while preserving the minimum-weight explanations of the retained syndrome. We formulate DEM contraction through a general mathematical framework applicable to arbitrary DEMs, together with an efficient specialization that exploits the structure of matchable decoding graphs.
Masking detector observations alone does not fully reduce decoding cost if the
decoder still operates on the original DEM, whose size and connectivity
strongly affect its complexity~\cite{Higgott2025sparseblossom,
FusionBlossom,wu2025minimumweightparityfactordecoder}. We therefore contract the DEM onto the retained detectors while preserving their minimum-weight explanations. We develop a general mathematical framework
for contraction over arbitrary DEMs, from which we derive an efficient
specialization for matchable decoding graphs.

Let $H_M$ and $H_{\overline{M}}$ denote the retained and masked-out rows of
$H$, respectively. Masking replaces the syndrome with $s=(s_M,0)$, yielding the
following decoding problems on the original and contracted DEMs:
\begin{equation}
\begin{aligned}
    e_M^*
    &=
    \underset{e}{\arg\min}\;
    W^\top e
    && \mathrm{s.t.}\quad
    H_Me=s_M,\;
    H_{\overline{M}}e=0,
    \\
    e_{\mathrm{red}}^*
    &=
    \underset{e_{\mathrm{red}}}{\arg\min}\;
    W_{\mathrm{red}}^\top e_{\mathrm{red}}
    && \mathrm{s.t.}\quad
    H_{\mathrm{red}}e_{\mathrm{red}}=s_M.
\end{aligned}
\end{equation}
For an exact contraction, their optimal explanations must have identical
weight and logical effect:
\begin{equation}
    W^\top e_M^*
    =
    W_{\mathrm{red}}^\top e_{\mathrm{red}}^*,
    \qquad
    Le_M^*
    =
    L_{\mathrm{red}}e_{\mathrm{red}}^*.
\end{equation}

To construct the reduced DEM, we shift from the \emph{decoding view} to a
\emph{fault-set view}. In the decoding view, $e$ is optimized to explain a
given retained syndrome $s_M$. In the fault-set view, the same vector denotes
a composite set of original faults selected by its nonzero entries. If
$e\in\ker H_{\overline{M}}$, their effects cancel on all masked-out detectors,
allowing the entire set to be collapsed into a single effective fault:
\begin{equation}
    e\in\ker H_{\overline{M}}
    \longmapsto
    f_{\mathrm{eff}}(e)
    \equiv
    (\sigma,\lambda,w)
    =
    \left(H_Me,\;Le,\;W^\top e\right),
\end{equation}
where $\sigma$, $\lambda$, $w$ are its retained-detector pattern, logical
effect, and weight. We construct the reduced fault set in four phases:

\paragraph{Phase I: Reduced-Fault-Set Initialization.} We first identify original faults that individually satisfy $H_{\overline{M}}e=0$ and can therefore be directly included in the reduced fault set without contraction. In a matchable decoding graph $G=(V, E)$, these correspond to original edges whose endpoints both lie in $V_M$ or terminate at the boundary $\partial$, and are directly retained in $\hat{E}$ (Alg.~\ref{alg:graph-contraction}, Line~\ref{line:Ehat}), as illustrated in Fig.~\ref{fig:graph contraction}(a).

\paragraph{Phase II: Hidden-Space Contraction.}
\label{paragraph:phase2}
We next contract composite fault sets $e\in\ker H_{\overline{M}}$, whose
effects cancel on all masked-out detectors. Since multiple sets may induce the
same retained pattern $\sigma$, we select its minimum-weight realization:
\begin{equation}
    e^*(\sigma)
    =
    \underset{\substack{
        e\in\ker H_{\overline{M}}\\
        H_Me=\sigma
    }}{\arg\min}
    \;W^\top e.
\end{equation}
The resulting $f_{\mathrm{eff}}(e^*)$, labeled by
$(\sigma,Le^*,W^\top e^*)$, is included in the reduced DEM. For a matchable DEM, $H$ is the binary incidence matrix of its decoding graph
$G$. A fault vector $e$ selects an edge-induced subgraph $G_e$, for which
\begin{equation}
\begin{aligned}
    (He)_v &= \deg_{G_e}(v)\pmod 2,\\
    H_{\overline M}e=0
    &\Longrightarrow
    \deg_{G_e}(v)=0\pmod 2,
    \quad \forall v\in V_{\overline M}.
\end{aligned}
\end{equation}
Thus, only retained vertices and the matching boundary $\partial$ can have odd
degree. By the handshaking lemma and trail-decomposition theorem, the odd-degree
vertices of each connected component can be paired by edge-disjoint open
trails~\cite{west2001introduction}. Splitting these trails at intermediate retained vertices preserves their
combined detector pattern, weight, and logical effect:
$\sigma_1\oplus\sigma_2=\sigma$,
$w_1+w_2=w$, and $\lambda_1\oplus\lambda_2=\lambda$.
The resulting segments have endpoints in $V_M\cup\{\partial\}$ and internal
vertices only in $V_{\overline M}$. Under non-negative fault weights, any
repeated vertex creates a closed subtrail whose removal cannot increase the
weight. It therefore suffices to retain the minimum-weight hidden-space path
$P^*(v,u)$ between each $v,u\in V_M\cup\{\partial\}$. Its effective edge has
weight $\sum_{e'\in P^*}w(e')$ and logical effect
$\bigoplus_{e'\in P^*}\ell(e')$. We generate them using Dijkstra's algorithm from each retained vertex,
traversing only masked-out vertices until reaching another retained vertex or
$\partial$ (Alg.~\ref{alg:graph-contraction}, Phase II;
Fig.~\ref{fig:graph contraction}(b)).

% \paragraph{Phase III: Weight Truncation.}
% The general contraction can contain up to $2^{r_M}-1$ non-trivial effective faults, where
% $r_M=\operatorname{rank}(H)-\operatorname{rank}(H_{\overline M})\le m_M$
% is the number of independent retained-syndrome degrees of freedom.
% To bound the reduced DEM size, we approximate the full contraction by
% discarding effective faults above a weight cutoff $\tau$, trading exact
% minimum-weight equivalence for sparsity. For a matchable DEM, the graph
% structure limits the candidate set to at most
% $\binom{m_M+1}{2}$ effective edges, which are further truncated as in
% Alg.~\ref{alg:graph-contraction} Line~\ref{line:truncate}, as illustrated in
% Fig.~\ref{fig:graph contraction}(c).
\paragraph{Phase III: Weight Truncation.}
An exact general contraction may contain up to $2^{r_M}-1$ non-trivial
effective faults, where
$r_M=\operatorname{rank}(H)-\operatorname{rank}(H_{\overline M})\leq m_M$
counts the independent retained-syndrome degrees of freedom. We therefore
discard effective faults with weight $w\geq\tau$, trading exact minimum-weight
equivalence for sparsity. For matchable DEMs, at most $\binom{m_M+1}{2}$ effective edges are generated
before truncation (Alg.~\ref{alg:graph-contraction},
Line~\ref{line:truncate}; Fig.~\ref{fig:graph contraction}(c)).

\paragraph{Phase IV: Feasibility Repair.}
Although weight truncation improves sparsity, it may remove effective faults
needed to represent valid retained syndromes. The full contraction spans exactly $\operatorname{col}(H_{\mathrm{red}})
    =
    \mathcal{S}_M
    =
    \left\{
        H_Me
        \mid
        H_{\overline M}e=0
    \right\}$, 
% \begin{equation}
%     \operatorname{col}(H_{\mathrm{red}})
%     =
%     \mathcal{S}_M
%     =
%     \left\{
%         H_Me
%         \mid
%         H_{\overline M}e=0
%     \right\},
% \end{equation}
whereas the truncated matrix $H_\tau$ may satisfy
$\operatorname{col}(H_\tau)\subsetneq\mathcal{S}_M$. Repair therefore restores
discarded faults until $\operatorname{col}(H_\tau)=\mathcal{S}_M$. For a matchable DEM, boundary reachability is sufficient for feasibility.
Consider a path
$v_i,v_{j_1},\ldots,v_{j_k},\partial$
from retained vertex $v_i$ to $\partial$. XORing its edge columns telescopes to
\begin{equation}
    (\mathbf e_i+\mathbf e_{j_1})
    \oplus\cdots\oplus
    (\mathbf e_{j_{k-1}}+\mathbf e_{j_k})
    \oplus\mathbf e_{j_k}
    =
    \mathbf e_i.
\end{equation}
Thus, boundary access makes the singleton syndrome $\mathbf e_i$
representable. Since Phase II preserves these paths through contraction,
boundary reachability of every retained vertex makes all singleton syndromes,
and hence all of $\mathbb F_2^{m_M}$, representable. After truncation, we identify connected regions without boundary access and
restore each region's minimum-weight discarded boundary edge, yielding the
final graph $G_M$ (Alg.~\ref{alg:graph-contraction}, Phase IV;
Fig.~\ref{fig:graph contraction}(d)).

\subsection{Two-Stage Early Escape}
\label{sec:magic firm sys:2stage}
\noindent\textbf{\textit{\# Observation:}}
\textit{The partial gap is a one-side biased approximation of the complete gap.}
For example, for $d_{\mathrm{cultiv}}=3$, $d_{\mathrm{escape}}=15$,
$p=10^{-3}$, and $K_Q=250$, we observe
$P(g_p=g_c)=0.734$, $P(g_p<g_c)=0.260$, and
$P(g_p>g_c)=0.006$. We therefore summarize their empirical relation as $g_p(s) \lesssim g_c(s)$. Although this suggests that $g_p$ could replace $g_c$ with a shifted threshold, the discrepancy is not uniform. Among the $26.0\%$ of
shots with $g_p<g_c$, many have a very small $g_p$ despite a large $g_c$,
consistent with the structural incompleteness captured by $S(M)$. Thus,
a conservative threshold rejects many acceptable attempts, while
lowering it compromises reliability. We therefore use $g_p$ only for
high-confidence early acceptance and fall back to $g_c$ otherwise, with two independently calibrated thresholds $t_h$ and $t_c$:
\begin{equation}
\operatorname{Accept}(s)=
\begin{cases}
1, & g_p(s)>t_h, \\[2pt]
\mathbf{1}\!\left[g_c(s)>t_c\right], & \text{otherwise}.
\end{cases}
\end{equation}
At runtime, two decoder pairs execute concurrently: one computes $g_p$
from the masked syndrome and contracted DEM, while the other performs
complete decoding. If $g_p>t_h$, the state advances immediately;
otherwise, the system waits for $g_c$. The complete decoder always
finishes in the background, with its correction tracked in the Pauli
frame for later logical measurements. Thus, early-accepted attempts
remove complete-gap evaluation from the state-availability critical
path, while all others retain the complete-only path.

\section{Evaluation}
\label{sec:evaluation}

\subsection{Configurable Control Runtime}
We compile each cultivation circuit into a control hierarchy and
per-board instructions. Across the circuit suite and monolithic and
distributed topologies, reconstructed detectors match Stim~\cite{Gidney_2021} on 1000
noisy shots per design with zero mismatches.
Multi-shot RTL tests verify the runtime control. We report control latency and FPGA results for six
representative circuits with \(d_{\mathrm{cultiv}}\in\{3,5\}\) and
\(d_{\mathrm{escape}}\in\{13,15,17\}\).
\subsubsection{Control Latency}
Each LCore has deterministic latency. Upward detector forwarding takes three cycles at a leaf, two at a router, and two to expose aggregated data at the root, while downward event relay takes one cycle per level. 
A \emph{stage LCore}, assigned to evaluate post-selection at a protocol stage,
generates a rejection in three cycles, and the root converts it into a global
\texttt{ABORT} in two cycles, or one cycle for root-local post-selection. Inter-board latency consists of a one-word transfer plus serialization of additional words. For a \(B\)-bit message, payload width \(W\), encoding overhead \(H\), and line rate \(R\), we model
\begin{equation}
    L_{\mathrm{link}}(B)=L_{\mathrm{1w}}+
\left(\left\lceil\frac{B}{W}\right\rceil-1\right)\frac{W+H}{R}.
\end{equation}
Following Liu et al.~\cite{liu2026scalableopensourceqecsubmicrosecond}, we use a 10-Gb/s GT link with 64b/66b encoding and conservatively set the bidirectional \(L_{\mathrm{1w}}\) to their larger measured one-way latency of $157ns$, giving $6.6 ns$ per additional word. Each parent--child pair uses a dedicated duplex link and carries at most one message per measurement round, so we omit queueing latency. For each circuit, we generate the control tree by recursively bisecting qubits along the longer spatial axis into leaves of at most \(Q\) qubits, then grouping their centroids with maximum fanout \(F\). Following \cite{liu2026scalableopensourceqecsubmicrosecond}, we set \(Q=14\) and \(F=29\). The compiler maps each detector and post-selection stage to the lowest board covering its measurements, with the decoder at the root. We model each compiled hierarchy in the cycle-accurate emulator using four critical-path latencies; accumulated across execution and retries, they form
the control component of the preparation-time breakdowns.
$L_{\mathrm{bcast}}$ is the root-to-leaf delivery latency for the initial
\texttt{START}. An attempt reaching gap decoding incurs $L_{\mathrm{del}}$
from leaf measurement availability to root decoder input and, after decoding,
$L_{\mathrm{gap}}$ from gap availability through event generation and
broadcast to all leaves. If post-selection at stage $s$ rejects earlier,
$L_{\mathrm{ps}}^{(s)}$ instead covers the path from its last required
measurement through the assigned stage LCore, root \texttt{ABORT} generation,
and broadcast to all leaves.

% We report four worst-case path latencies. \(L_{\mathrm{del}}\) spans leaf measurement availability to root decoder input, while \(L_{\mathrm{bcast}}\) spans root event issuance to all leaves. \(L_{\mathrm{gap}}\) spans \(g_c/g_p\) availability at the root to the corresponding event at all leaves, excluding decoder execution.
% \(L_{\mathrm{ps}}^{(s)}\) spans the last measurement of stage \(s\) becoming available at a leaf to the resulting ABORT reaching all leaves.
\begin{table}[t]
\centering
\small
\setlength{\tabcolsep}{3pt}
\caption{\textbf{Compiled control configurations and worst-case latencies.}
Tree lists leaf/router/root boards. All latencies are in $ns$, and
$L_{\mathrm{ps}}^{\max}$ is maximized across stages. LCore clock frequency set to 100\,MHz}
\label{tab:control-latency}
\begin{tabular}{@{}crcrrr@{}}
\toprule
$d_{cultiv}/d_{escape}$ & Qubits & Tree (L/R/Rt)
& $L_{\mathrm{del}}$
& $L_{\mathrm{gap}}$
& $L_{\mathrm{ps}}^{\max}$ \\
\midrule
$3/13$ & 342 & $32/2/1$ & 410.4 & 344.0 & 764.4 \\
$3/15$ & 454 & $38/2/1$ & 417.0 & 344.0 & 771.0 \\
$3/17$ & 582 & $64/4/1$ & 403.8 & 344.0 & 757.8 \\
$5/13$ & 351 & $32/2/1$ & 410.4 & 344.0 & 764.4 \\
$5/15$ & 463 & $47/2/1$ & 423.6 & 344.0 & 777.6 \\
$5/17$ & 591 & $64/4/1$ & 410.4 & 344.0 & 764.4 \\
\bottomrule
\end{tabular}
\end{table}
Table~\ref{tab:control-latency} shows that all six configurations form
three-level trees containing 35--69 boards, with maximum uplink messages
of 217--371 bits (4--6 words). Because their tree depths are identical,
$L_{\mathrm{bcast}}$ remains 334\,ns and $L_{\mathrm{gap}}$ remains
344\,ns. Syndrome delivery varies from 403.8 to 423.6\,ns, while
post-selection feedback remains below 0.78\,$\mu$s across all
configurations.
\subsubsection{FPGA Implementation}
We evaluate the compiled LCore designs on an AMD
Virtex UltraScale+ VU19P using Vivado 2022.1.
Table~\ref{tab:fpga-resource} reports the maximum per-board resource
usage and minimum \(F_{\max}\) for each board role across all six
configurations. Even the largest LCore uses less than \(0.3\%\) of the VU19P LUTs,
indicating a lightweight footprint for integration into existing
FPGA-based control systems.
\begin{table}[t]
    \centering
    \small
    \setlength{\tabcolsep}{3pt}
    \caption{Post-route LCore results on the VU19P.
    Resources are per-board maxima, whereas $F_{\max}$ is the minimum
    within each role. Header denominators give device capacities; BRAM is
    reported in 36-Kb tile equivalents.}
    \label{tab:fpga-resource}
    \begin{tabular}{@{}lrrrrr@{}}
        \toprule
        Role
        & \shortstack{LUT\\(/4,085,760)}
        & \shortstack{FF\\(/8,171,520)}
        & \shortstack{BRAM\\(/2,160)}
        & \shortstack{DSP\\(/3,840)}
        & \shortstack{$F_{\max}$\\(MHz)} \\
        \midrule
        Leaf   & 846    & 658   & 0.5 & 0 & $\geq 500$ \\
        Router & 3,888  & 3,667 & 0   & 0 & 404 \\
        Root   & 11,624 & 8,051 & 0   & 0 & 331 \\
        \bottomrule
    \end{tabular}
\end{table}

% \begin{table}[t]
%     \centering
%     \small
%     \setlength{\tabcolsep}{4pt}
%     \caption{Post-route LCore results on the VU19P.
%     Resources are per-board maxima, whereas \(F_{\max}\) is the minimum
%     within each role. BRAM is reported in 36-Kb tile equivalents.}
%     \label{tab:fpga-resource}
%     \begin{tabular}{@{}lrrrrr@{}}
%         \toprule
%         Role & LUT & FF & BRAM & URAM & \(F_{\max}\) (MHz) \\
%         \midrule
%         Leaf   & 846    & 658   & 0.5 & NA & \(\geq 500\) \\
%         Router & 3,888  & 3,667 & NA   & NA & 404 \\
%         Root   & 11,624 & 8,051 & NA   & NA & 331 \\
%         \bottomrule
%     \end{tabular}
% \end{table}

\subsection{Partial-Gap Estimation Ablation}
We evaluate partial-gap estimation in three steps: comparing mask-construction methods at matched size, sweeping retained-detector coverage, and isolating the effect of DEM contraction on gap preservation and decoding cost.
\begin{table}[htbp]
    \centering
    \caption{Mask-construction comparison at matched sizes.
    $P_{\circ}\equiv P(g_p\circ g_c)$ for
    $\circ\in\{=,>,<\}$; lower $S(M)$ indicates less
    boundary-straddling fault mass.}
    \label{tab:mask-construction}

    \footnotesize
    \setlength{\tabcolsep}{3.5pt}
    \renewcommand{\arraystretch}{1.08}

    \begin{tabular}{@{}c l c c c c c@{}}
        \toprule
        $d_{\mathrm{esc}}$ & Strategy
        & $|M|$ & $S(M)$ & $P_{=}$ & $P_{>}$ & $P_{<}$ \\
        \midrule

        \multirow{3}{*}{13}
        & Geometric     & 335 & 1.70 & 0.567 & 0.063 & 0.370 \\
        & LAP           & 336 & 2.01 & 0.598 & 0.006 & 0.396 \\
        & LAP + closure & 328 & \textbf{1.17} & \textbf{0.739}
                                      & \textbf{0.006} & \textbf{0.255} \\
        \midrule

        \multirow{3}{*}{15}
        & Geometric     & 335 & 2.14 & 0.501 & 0.166 & 0.334 \\
        & LAP           & 336 & 2.31 & 0.566 & 0.007 & 0.427 \\
        & LAP + closure & 336 & \textbf{1.18} & \textbf{0.734}
                                      & \textbf{0.006} & \textbf{0.260} \\
        \midrule

        \multirow{3}{*}{17}
        & Geometric     & 336 & 1.80 & 0.502 & 0.034 & 0.464 \\
        & LAP           & 336 & 2.45 & 0.562 & 0.007 & 0.431 \\
        & LAP + closure & 335 & \textbf{1.02} & \textbf{0.759}
                                      & \textbf{0.007} & \textbf{0.234} \\
        \bottomrule
    \end{tabular}
\end{table}
\subsubsection{Mask-Construction Strategy}
We isolate the effects of LAP and Structural-Closure using \(d_{\mathrm{cultiv}}=3\), \(p=10^{-3}\), and \(d_{\mathrm{escape}}\in\{13,15,17\}\). At matched sizes of 328--336 detectors, we compare a handcrafted geometric mask, LAP, and LAP with closure. We evaluate each mask on \(2\times10^5\) post-selected shots using the complete DEM to exclude contraction effects. As shown in Table~\ref{tab:mask-construction}, although geometric masks have lower $S(M)$ than LAP alone, they exhibit much larger high-side mismatch: $P_{>}=3.4\%$--$16.6\%$, compared with $0.6\%$--$0.7\%$ for LAP. This isolates LAP's main contribution: suppressing $g_p$ overestimation by retaining logically influential detectors, while improving exact agreement by 3.1--6.5 percentage points. Closure then reduces $S(M)$ by 42--58\% over LAP and raises agreement by another 14.1--19.7 points while retaining $P_{>}\leq0.7\%$. Together, LAP and closure improve both structural quality and gap fidelity over simplest heuristic geometric selection.

\subsubsection{Mask-Size Sensitivity}
\begin{figure}[htbp]
    \centering
    \includegraphics[width=0.75\linewidth]{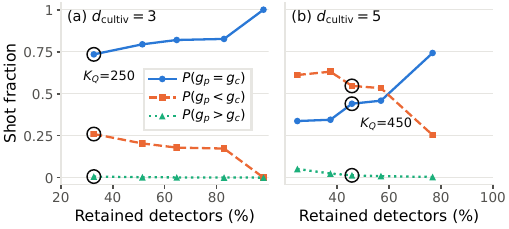}
    \caption{\textbf{Mask-size sensitivity} for
    $d_{\mathrm{escape}}=15$ and $p=10^{-3}$. Each point uses
    $2\times10^5$ post-selected shots; black circles mark the masks used in
    the end-to-end evaluation.}
    \label{fig:mask size}
\end{figure}

% We next vary the LAP-with-closure mask size for
% \(d_{\mathrm{cultiv}}\in{3,5}\). Because closure changes the final
% mask size, \figref{fig:mask size} reports the actual retained-detector
% fraction \(\lvert M\rvert/\lvert D\rvert\), rather than \(K_Q\).
% \begin{figure}[htbp]
%     \centering
%     \includegraphics[width=\linewidth]{figures/mask_size.pdf}
%     \caption{\textbf{Mask-size sensitivity} for \(d_{\mathrm{escape}}=15\) and
% \(p=10^{-3}\). Each point uses \(2\times10^5\) post-selected shots;
% black circles mark the masks used in the later end-to-end evaluation.
% }
%     \label{fig:mask size}
% \end{figure}
% Larger masks generally increase $P_{=}$ and reduce $P_{>}$. At the shared $K_Q=250$, $d_{\mathrm{cultiv}}=3$ achieves $P_{=}=73.4\%$ and $P_{>}=0.6\%$, whereas $d_{\mathrm{cultiv}}=5$ reaches only $33.7\%$ and $5.2\%$. The larger code spreads logical ambiguity across a broader detector region, making its complete gap harder to preserve with a compact mask. Nevertheless, $P_{>}$ remains far below $P_{<}$ across practical mask sizes, preserving the conservative bias for early acceptance. The selected $K_Q=250/450$ masks retain $32.7\%/45.7\%$ of detectors.
We vary the LAP-with-closure mask size for
$d_{\mathrm{cultiv}}\in\{3,5\}$. Since closure alters the final size,
\figref{fig:mask size} reports the actual retained-detector fraction
$|M|/|D|$ rather than $K_Q$.
Larger masks generally raise $P_{=}$ and reduce $P_{>}$. At the shared
$K_Q=250$, $d_{\mathrm{cultiv}}=3$ achieves $P_{=}=73.4\%$ and
$P_{>}=0.6\%$, whereas $d_{\mathrm{cultiv}}=5$ reaches only $33.7\%$ and
$5.2\%$. The larger code spreads logical ambiguity over a broader detector
region, making the complete gap harder to preserve with a compact mask. Nevertheless,
$P_{>}$ remains far below $P_{<}$ across practical mask sizes, preserving the
conservative bias for early acceptance. The selected $K_Q=250/450$ masks
retain $32.7\%/45.7\%$ of detectors.

\subsubsection{Effect of DEM Contraction}
% \begin{figure*}
%     \centering
%     \includegraphics[width=515.52pt]{figures/contraction_pair_d1-3.pdf} % 0.95\linewidth]
%     \caption{\textbf{MagiC-Firm control-system architecture.} The left panels show offline compilation and hierarchical runtime deployment over an existing physical-control substrate, where PCore denotes the physical control core that directly controls the underlying qubits; the center panels detail the LCore and MMIO architectures; and the right panel expands the configurable detector-construction datapath. Gray components denote the existing physical-control system, while colored blocks and black interconnects denote MagiC-Firm extensions.}
%     \label{fig:control system arch}
%     \Description{}
% \end{figure*}
\begin{figure}[htbp]
    \centering
    \includegraphics[width=\linewidth]{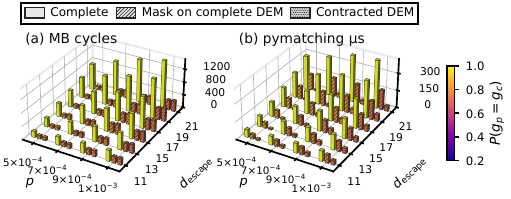}
    \caption{\textbf{DEM-contraction ablation} over 24
\(d_{\mathrm{cultiv}}=3\) configurations. Bar height reports the slower
of the two parallel complementary-gap decodes, measured in (a) Micro
Blossom~\cite{10.1145/3676641.3716005} cycles and (b) PyMatching~\cite{Higgott2025sparseblossom} microseconds; color denotes
\(P(g_p=g_c)\), with unity for complete decoding.
}
    \label{fig:dem-contraction}
\end{figure}
We compare three decoding setups on identical shots to separate the
effects of masking and contraction: the complete syndrome on the
complete DEM, the masked syndrome on the complete DEM, and the masked
syndrome on the contracted DEM. We evaluate 24
$d_{\mathrm{cultiv}}=3$ configurations spanning
odd $d_{\mathrm{escape}}$ from 11 to 21 and
$p\in\{5,7,9,10\}\times10^{-4}$, using a separately selected mask for
each circuit ($K_Q=100$--$300$, $|M|=144$--$442$) and a contraction
cutoff of 15 (equivalent fault probability $\approx3.1\times10^{-7}$). \figref{fig:dem-contraction} shows that contraction leaves the partial
gap nearly unchanged. 
% Across $2\times10^5$ post-selected shots per
% configuration, the median and maximum changes in $P(g_p=g_c)$ relative
% to the masked/full-DEM baseline are only 0.007 and 0.55 percentage
% points. More importantly for early acceptance, $P(g_p>g_c)$ changes by
% at most 0.016 points and remains below 1.1\%, showing negligible
% additional high-side mismatch. 
Across \(2\times10^5\) post-selected shots per configuration, replacing the complete DEM with the contracted DEM changes \(P(g_p=g_c)\) by a median of only \(0.007\) percentage points (pp), with a worst-case change of \(0.55\) pp. More importantly for early acceptance, it changes \(P(g_p>g_c)\) by at most \(0.016\) pp, while remains below \(1.1\%\) across all configurations, showing negligible additional high-side mismatch.
In \figref{fig:dem-contraction}(b), masking reduces PyMatching latency
by $47.5\%$ at the median, and contraction cuts a further $68.4\%$, yielding
$83.9\%$ overall. By contrast, masking already captures most of the cycle reduction for
spatially mapped Micro Blossom in \figref{fig:dem-contraction}(a), both contraction and original DEM remain around $66.3\%$
below complete decoding. Its main hardware benefit is instead avoiding full duplication of the complete decoder's graph-dependent resources when adding the partial path. Because
Micro Blossom instantiates processing units for graph vertices and
edges~\cite{10.1145/3676641.3716005}, contraction reduces this incremental
vertex- and edge-unit footprint by medians of $73.1\%$ and $60.4\%$.
It also lowers the partial graph's maximum detector degree by $82.4\%$,
easing routing around the observable detector and potentially improving
$F_{\max}$.
% Contraction increases cycles by $8.4\%$ at the median, yet remains $66.3\%$
% below complete decoding.
% Its main hardware benefit is instead
% scalability: because Micro Blossom instantiates a processing unit per
% graph vertex and edge~\cite{10.1145/3676641.3716005}, contraction reduces
% detector vertices and fault edges by medians of $73.1\%$ and $60.4\%$.
% Thus, the two parallel instances required for complementary-gap decoding still use fewer processing units than one complete decoder.
% Contraction also lowers the maximum detector degree by $82.4\%$, easing routing around the observable detector and potentially improving \(F_{\max}\).

\subsection{End-to-End Two-Stage Early Escape Evaluation}
We evaluate the mean wall-clock preparation time per accepted state, including failed attempts. Our cycle-accurate simulator accounts for quantum execution, control feedback, and RTL-characterized Micro Blossom latency; LER is measured independently from large-sample gap tables. Both decoders use the distance-matched \(F_{\max}\) reported by Micro Blossom, with its \(d=15\) value used for larger \(d_{\mathrm{escape}}\)~\cite{10.1145/3676641.3716005}. This is conservative: the reported \(F_{\max}\) is characterized on standard surface-code workloads, whereas our more complex cultivation graph may clock lower; assigning the same clock to the smaller, lower-degree partial graph also ignores its potential frequency advantage.
\begin{figure}[htbp]
    \centering
    \includegraphics[width=0.8\linewidth]{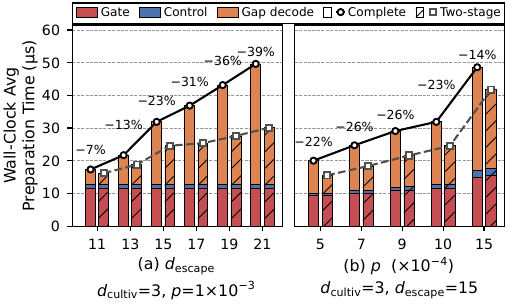}
    \caption{\textbf{End-to-end preparation time at matched LER} for \(d_{\mathrm{cultiv}}=3\): (a) scaling with \(d_{\mathrm{escape}}\) at \(p=10^{-3}\), and (b) scaling with \(p\) at \(d_{\mathrm{escape}}=15\). Stacked bars decompose gate execution, control, and gap decoding; annotations report the reduction in total preparation time.
}
    \label{fig:iso ler scaling}
\end{figure}
\subsubsection{Iso-LER Scaling}
We use disjoint selection and evaluation sets: at \(t_c=35\), the former selects the smallest \(t_h\) whose LER is at most \(25\%\) above complete-only decoding, while the latter supplies reported LERs. Each two-stage point is matched by a randomized per-preparation mixture of bracketing complete thresholds. In \figref{fig:iso ler scaling}(a), savings increase with \(d_{\mathrm{escape}}\), reaching \(39.3\%\). This trend arises because logical ambiguity remains concentrated around the fixed \(d_{\mathrm{cultiv}}=3\) region, so the mask grows slower than the complete DEM; its retained-detector fraction falls from \(52.9\%\) to \(17.5\%\). \figref{fig:iso ler scaling}(b) shows \(22\%\)--\(26\%\) savings through \(p=10^{-3}\), falling to \(14.1\%\) at the highest \(p\) as retry and residual-decoding costs grow. Nearly unchanged gate and control time across both sweeps confirms that gap-decoding savings translate directly into system speedup rather than hiding additional retries.
\subsubsection{Design-space Exploration and Sensitivity Analysis}
\begin{figure}[htbp]
    \centering
    \includegraphics[width=0.8\linewidth]{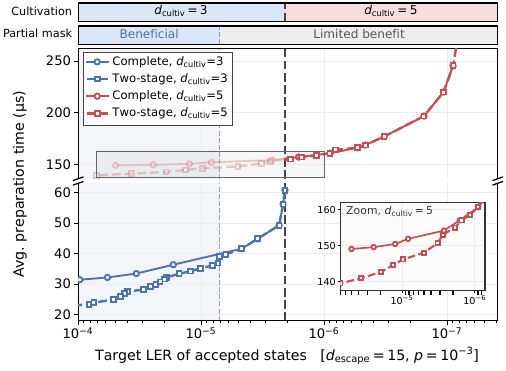}
    \caption{\textbf{Preparation-time--LER Pareto frontiers} at \(d_{\mathrm{escape}}=15\) and \(p=10^{-3}\). Top bands summarize the preferred \(d_{\mathrm{cultiv}}\) and the region benefiting from partial decoding. The faded $d_{\mathrm{cultiv}}=5$ segment is dominated by
$d_{\mathrm{cultiv}}=3$, which achieves lower preparation time at matched LER.
}
    \label{fig:iso dse}
\end{figure}

\begin{figure}[htbp]
    \centering
    \includegraphics[width=0.8\linewidth]{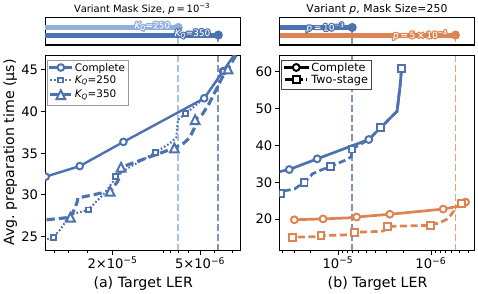}
    \caption{\textbf{Sensitivity of the two-stage frontier.} (a) variant $K_Q$ and (b) variant $p$, for \(d_{\mathrm{cultiv}}=3\), \(d_{\mathrm{escape}}=15\). Vertical lines mark the lowest LER retaining a two-stage advantage.
}
    \label{fig:sensitivity}
\end{figure}

We sweep \(t_c\) and all two-stage pairs \((t_c,t_h)\) at \(d_{\mathrm{escape}}=15\) and \(p=10^{-3}\) to construct held-out Pareto frontiers. Since complete decoding is a limiting case of the two-stage policy, the optimized two-stage frontier can never be worse than the complete frontier. In \figref{fig:iso dse}, \(d_{\mathrm{cultiv}}=3\) benefits down to approximately \(6\times10^{-6}\), below which the curves merge. Although \(d_{\mathrm{cultiv}}=5\) reaches lower LERs, its broader logical ambiguity favors larger masks and reduces the early-accept advantage. For \(d_{\mathrm{cultiv}}=3\), \figref{fig:sensitivity}(a) shows that increasing \(K_Q\) modestly extends this operating region at the cost of longer partial decoding. By contrast, halving \(p\) at fixed \(K_Q=250\) lowers the beneficial-range boundary by over an order of magnitude, from \(7.1\times10^{-6}\) to \(5.5\times10^{-7}\) [\figref{fig:sensitivity}(b)]. Thus, the operating region scales favorably with improving physical fidelity, while mask size provides a finer trade-off between latency and low-LER reach.
\subsubsection{Potential System-Level Impact}
Based on the operating region at \(p=10^{-3}\), partial gap is particularly well suited to cultivation used as an intermediate source for subsequent distillation, as discussed in \secref{sec:background:msc:usecase}. Representative large-scale FTQC may target magic-state LERs with \(\mathrm{LER_{magic}}\sim10^{-12}\)--\(10^{-13}\)~\cite{beverland2022assessingrequirementsscalepractical}; under the 15-to-1 approximation \(\mathrm{LER_{magic}}\approx35\mathrm{LER_{cultiv}}^3\), this permits \(\mathrm{LER_{cultiv}}\in[1.4,\,3.1]\times10^{-5}\). Replacing Litinski's~\cite{Litinski_2019} first-level blocks with \(N\) parallel cultivation sources gives $T_{\mathrm{magic}}(N)=7.5\max(
t_{\mathrm{rot}},\frac{2T_{\mathrm{cultiv}}}{N})$. At \(\mathrm{LER_{cultiv}}=2\times10^{-5}\), with \(t_{\mathrm{rot}}=15\mu\mathrm{s}\)($d$=15 surface code) and \(N=2\), partial gap reduces \(T_{\mathrm{magic}}\) and the source capacity required at fixed throughput by approximately \(12\%\), or equivalently increases throughput by \(13.6\%\). Because magic-state supply could dominate the resources and runtime of large-scale FTQC systems~\cite{9251988,Goings_2022,campbell2026resourceestimationefficientcompilation}, a representative \(90\%\) source-bound workload would see roughly an \(11\%\) reduction in application runtime. This illustrates a practical regime in which partial gap can potentially translate its fidelity--latency trade-off into system-level benefits.
\section{Future Work}
\label{sec:future work}
MagiCFirm currently targets statically compiled circuits, with runtime events selecting among precompiled control paths. 
Supporting dynamic circuits, such as adaptive syndrome extraction~\cite{Berthusen_2025}, will require richer data-dependent event triggers and conditional microprogram branches.
Real systems must also accommodate changing device conditions and fault information.
Online DEM updates~\cite{ziad2026greenpeasunlockingadaptivequantum} could track such drift, while per-shot partial-mask adaptation could exploit leakage and erasure information. 
Finally, cultivation is only a starting point: the same detector-driven, event-controlled runtime can be extended to other post-selection-based FTQC protocols, moving toward a reusable execution substrate for probabilistic FTQC.
% MagiCFirm currently targets statically QEC schedules, with events
% selecting precompiled paths. Supporting Dynamic QEC~\cite{eickbusch2025demonstratingdynamicsurfacecodes}, like adaptive syndrome extraction~\cite{Berthusen_2025},
% will require data-dependent triggers and conditional microprogram branches.
% Online DEM updates~\cite{ziad2026greenpeasunlockingadaptivequantum} could track calibration drift, while per-shot mask
% adaptation could exploit leakage and erasure information. Beyond cultivation,
% the same detector-driven, event-controlled runtime could support other
% post-selection-based FTQC protocols, advancing toward a reusable execution
% substrate for early-stage FTQC.
\section{Conclusion}
\label{sec:conclusion}
% We presented MagiCFirm, a configurable hardware runtime for end-to-end
% magic-state cultivation. Offline-compiled microprograms drive hierarchical
% event control and detector construction, while profiled partial masking and
% DEM contraction enable two-stage early escape alongside complete decoding.
% Evaluation shows low-latency control, a lightweight FPGA footprint, and
% reduced preparation time at matched LER. MagiCFirm thus turns cultivation
% from a coarse logical abstraction into a latency- and resource-accountable
% runtime operation.
We presented MagiCFirm, a configurable hardware runtime for end-to-end
magic-state cultivation. It combines offline-compiled microprograms with
hierarchical event handling and configurable detector construction. Its
partial-decoding flow uses Logical Ambiguity Profiling, Structural Closure, and
DEM contraction to construct a compact decoding problem, enabling two-stage
early escape alongside complete decoding. Evaluation demonstrates low-latency
control with a lightweight FPGA footprint and reduced preparation time at
matched LER. Together, these results turn cultivation from a coarse logical
abstraction into a concrete, latency- and resource-accountable runtime
operation.

% MagiCFirm currently targets statically compiled QEC schedules, with runtime
% events selecting among precompiled control paths. Supporting dynamic QEC, such
% as adaptive syndrome extraction, will require richer data-dependent event
% triggers and conditional microprogram branches. Real systems must also
% accommodate changing device conditions and fault information. Online DEM
% updates could track calibration drift, while per-shot DEM and partial-mask
% adaptation could exploit leakage and erasure information. Finally, cultivation
% is only a starting point: the same detector-driven, event-controlled runtime
% can be extended to other post-selection-based FTQC protocols, moving toward a
% reusable execution substrate for probabilistic FTQC.

\clearpage

\bibliographystyle{ACM-Reference-Format}
\bibliography{references}

\end{document}